\documentclass[fleqn,usenatbib]{mnras}

\usepackage{newtxtext,newtxmath}

\usepackage[T1]{fontenc}

\DeclareRobustCommand{\VAN}[3]{#2}
\let\VANthebibliography\thebibliography
\def\thebibliography{\DeclareRobustCommand{\VAN}[3]{##3}\VANthebibliography}

\usepackage{graphicx}	
\usepackage{amsmath}	
\usepackage{amssymb}	
\usepackage[caption=false]{subfig}
\usepackage[normalem]{ulem}

\newcommand{\popp}{\boldsymbol{\Psi}}

\newcommand{\data}{\mathbf{d}}
\newcommand{\pr}{\text{Pr}}
\newcommand{\de}{\text{d}}
\newcommand{\kpc}{\text{kpc}}
\newcommand{\Msun}{M_{\odot}}
\newcommand{\kmsec}{\text{km}/\text{s}}
\newcommand{\pc}{\text{pc}} 
\newcommand{\yr}{\text{yr}} 
\newcommand{\Msunppcc}{\Msun\pc^{-3}}
\newcommand{\Msunppcsquare}{\Msun\pc^{-2}}

\newcommand{\mas}{\text{mas}}
\newcommand{\arcs}{\text{arcs}}

\newcommand{\Xs}{\tilde{X}}
\newcommand{\Ys}{\tilde{Y}}
\newcommand{\Zs}{\tilde{Z}}
\newcommand{\Us}{\tilde{U}}
\newcommand{\Vs}{\tilde{V}}
\newcommand{\Ws}{\tilde{W}}

\title[Stellar Streams and the Disk Matter Density]{Measuring the Matter Density of the Galactic Disk Using Stellar Streams}

\author[A. Widmark et al.]{
Axel Widmark,$^{1,2}$\thanks{E-mail: axel.widmark@fysik.su.se}
Khyati Malhan,$^{1,2}$
Pablo F. de Salas$^{1,2}$
Sofia Sivertsson$^{1,2}$
\\
$^{1}$Department of Physics, AlbaNova, Stockholm University, SE-106 91, Stockholm, Sweden\\
$^{2}$The Oskar Klein Centre for Cosmoparticle Physics, AlbaNova, SE-106 91 Stockholm, Sweden\\
}

\date{Accepted June 9. Received June 9; in original form March 16}

\pubyear{2020}

\hypersetup{draft}
\begin{document}
\label{firstpage}
\pagerange{\pageref{firstpage}--\pageref{lastpage}}
\maketitle

\begin{abstract}
We present a novel method for determining the total matter surface density of the Galactic disk by analysing the kinematics of a dynamically cold stellar stream that passes through or close to the Galactic plane. The method relies on the fact that the vertical component of energy for such stream stars is approximately constant, such that their vertical positions and vertical velocities are interrelated via the matter density of the Galactic disk. By testing our method on mock data stellar streams, with realistic phase-space dispersions and \emph{Gaia} uncertainties, we demonstrate that it is applicable to small streams out to a distance of a few kilo-parsec, and that the surface density of the disk can be determined to a precision of $6~\%$. This method is complementary to other mass measurements. In particular, it does not rely on any equilibrium assumption for stars in the Galactic disk, and also makes it possible to measure the surface density to good precision at large distances from the Sun. Such measurements would inform us of the matter composition of the Galactic disk and its spatial variation, place stronger constraints on dark disk sub-structure, and even diagnose possible non-equilibrium effects that bias other types of dynamical mass measurements.
\end{abstract}




\section{Introduction}

Determining the total matter density of the Galactic disk is of great importance for constraining the composition and dynamics of the Galaxy \citep{1998MNRAS.294..429D, Klypin:2001xu, Widrow:2008yg, 2010A&A...509A..25W,McMillan:2011wd,2014ApJ...794...59K,2017MNRAS.465...76M,2019arXiv190905269N,2019arXiv191104557C,2019arXiv191202086L}, has implications for direct and indirect dark matter detection experiments \citep{Jungman:1995df,2015PrPNP..85....1K}, and can inform us about the possibility of dark substructures \citep{10.1111/j.1365-2966.2008.13643.x,0004-637X-703-2-2275,Fan:2013tia,2014MNRAS.444..515R}. 

Usually, dynamical mass measurements are performed by isolating a stellar tracer population and fitting a function to its phase-space distribution \citep{Kapteyn1922, Oort1932, 1984ApJ...276..169B, 1984ApJ...287..926B, KuijkenGilmore1989a,KuijkenGilmore1989b,KuijkenGilmore1989c,KuijkenGilmore1991,Creze1998,HolmbergFlynn2000,Bienayme:2005py,doi:10.1111/j.1365-2966.2012.21608.x,0004-637X-756-1-89,0004-637X-772-2-108}. Under an assumption of equilibrium, the tracer population's velocity and number density distributions are interrelated via the gravitational potential. Although many studies on the local matter density of the Galactic disk quote rather small statistical uncertainties \citep{Read2014, Sivertsson:2017rkp, Schutz:2017tfp, Buch:2018qdr, 2019MNRAS.482..262W, Benito:2019ngh,2019A&A...623A..30W, Karukes:2019jxv,deSalas:2019pee}, there are significant disagreements between studies, and even between different tracer population samples within the same study. These discrepancies could potentially be due to data systematics, modelling differences, or non-equilibrium effects. With \emph{Gaia}'s second data release \citep{2016A&A...595A...1G,2018A&A...616A...1G}, it has become all the more clear that the Galaxy is perturbed, potentially by bar \citep{2018MNRAS.481.3794H,Khoperskov2019}, spiral structures \citep{2017A&A...597A..39M,2018MNRAS.481.1501B}, satellites and past mergers \citep{Antoja_2018,2019ApJ...874....3N, 2019MNRAS.482.1417B,2019A&A...621A..48L,lopezcorredoira2020gaiadr2}. It is an open question how and to what extent such non-equilibrium effects can bias dynamical mass measurements of our Galaxy.

In this work, we examine an alternative way of inferring the matter density of the Galactic disk through the use of stellar streams. A stellar stream is formed by the tidal stripping of a stellar body, such as a globular cluster or a dwarf galaxy, as it orbits the gravitational potential of its host galaxy \citep{Dehnen2004}. A key feature of stream structures is that their locus approximates an orbit, rendering the distribution of stream stars tightly constrained in phase-space. This property of streams is often exploited in order to measure the gravitational potential of the Milky Way and probe the dark matter distribution of the Galactic halo \citep{Koposov2010,2010ApJ...714..229L,Bovy2016GD1Pal5, 2019MNRAS.486.2995M}. These measurements are generally achieved by analysing streams with intermediate to large galactocentric radii (see \citealt{GrillmairCarlin2016} and references therein for a list of known stellar streams).

Our aim is to test if low-mass, dynamically cold stellar streams passing through or close to the Galactic plane can provide useful constraints to the matter density of the Galactic disk. This method is highly complementary to using a tracer population of the disk itself, as it does not rely on any equilibrium assumptions for disk stars. Neither does it depend on the stellar number density distribution, making it robust with respect to uncontrolled incompleteness effects. Our method accounts for the full error covariance matrix for all stream stars, as well as uncertainties associated with the Galactic potential and solar velocities. Our method of inference is local and does not rely on assumptions about the global structure of the Milky Way; furthermore, the stellar stream's phase-space dispersion is fitted in a data-driven manner. Hence, our method is relatively model independent, in comparison to many other common stream fitting methods which rely on stream simulation in a global Milky Way potential. We demonstrate that for a small stream of no more than 300 observed stars, we can constrain the surface density within 240 pc of the Galactic mid-plane to within an uncertainty of $6~\%$. This level of precision is competitive with other methods using disk tracer stars.

This paper is organised as follows. In Section~\ref{sec:gal_model}, we present the Galactic model used to simulate the streams and define our system of coordinates. In Section~\ref{sec:perf_orbit}, we demonstrate the principles and some limitations to our method by applying it on an ideal stream. This is followed by a description of our full model, applicable to realistic streams, described in Section~\ref{sec:model}. In Section~\ref{sec:sim_streams}, we describe how our mock data stellar streams are generated, and in Section~\ref{sec:results} we present our inferred results. Finally, we conclude in Section~\ref{sec:discussion}. Some technical aspects of our method is described in detail in Appendix~\ref{app:posterior}.

\section{Galactic model and coordinate system}\label{sec:gal_model}

For the purposes of simulating mock data stellar streams, we adopt the Milky Way model from \citet{2015ApJS..216...29B} and use the \texttt{gala} package \citep{gala}. This mass model of the Galaxy consists of a Hernquist bulge and nucleus, a Miyamoto-Nagai disk and an NFW halo \citep{1990ApJ...356..359H,1975PASJ...27..533M, 1997ApJ...490..493N}.

Throughout the paper, we use a solar rest frame coordinate system with spatial coordinates $\boldsymbol{X} = \{X,Y,Z\}$, where the spatial origin ($\boldsymbol{X} = 0$) is located at the Sun's position. The direction of positive $X$ is towards the Galactic centre, positive $Y$ is in the direction of the Galaxy's rotation and positive $Z$ is in the direction of the Galactic north, giving a right-handed coordinate system. Their respective time-derivatives give the velocities $\boldsymbol{V} = \{U,V,W\} = \{\dot{X},\dot{Y},\dot{Z}\}$, whose origin is the Sun's velocity. In our model, the Sun is located 15 pc above the Galactic mid-plane \citep{10.1093/mnras/stx729,2019MNRAS.482..262W}, and has a velocity of $\boldsymbol{V}_\odot = \{11,\, 12+232,\, 7.2\}$ km/s with respect to the Galaxy, corresponding to the Sun's peculiar motion \citep{10.1111/j.1365-2966.2010.16253.x} plus the rotational velocity at the solar radius of our Galactic model \citep{2019ApJ...871..120E}.

We also define an additional ``stream frame'' coordinate system, in which the phase-space distribution of a stellar stream will be expressed. The spatial coordinates are written as $\boldsymbol{\Xs} = \{\Xs,\Ys,\Zs\}$, and are analogous to $\boldsymbol{X}$; the direction of positive $\Zs$ is that of Galactic north, but the directions of positive $\Xs$ and $\Ys$ correspond to the directions of the Galactic centre and disk rotation as seen from the position of the stream. The origin of $\boldsymbol{\Xs}$ corresponds to the position where the stream intersects the Galactic mid-plane, and is static in the Galactic rest frame. Their respective time-derivatives give the velocities $\boldsymbol{\Vs} = \{\Us,\Vs,\Ws\}$, whose origin corresponds to that of the Galactic rest frame.

While we use a global potential when generating the mock data stellar streams, our model of statistical inference is local, and independent of assumptions about the global structure of the Galaxy. In this work, we consider a nearby stream passing through the stellar disk, and we limit ourselves to a distance of $\sim 400~\pc$ from the Galactic mid-plane. The gravitational potential in our model of inference is assumed to be locally axisymmetric, and separable in the vertical ($\Zs$) and horizontal ($\Xs$ and $\Ys$) directions. Close to the Galactic disk, it can be approximated as
\begin{equation}\label{eq:potential}
    \Phi(\boldsymbol{\Xs}) = 
    \Phi_{\Zs}(\Zs) - F_{\Xs} \Xs,
\end{equation}
where $F_{\Xs} = -\partial \Phi/\partial \Xs $ is a constant force per mass in the direction of the Galactic centre. While separability in the vertical direction is not an actual property of the Galactic potential, it is a valid approximation close to the Galactic disk. This approximation is tested in Section~\ref{sec:perf_orbit}, using our own method. Furthermore, simulations \citep{Garbari2011} and Milky Way observations \citep{10.1093/mnras/stv1314,Sivertsson:2017rkp} indicate that separability is a valid assumption to at least within 500 pc from the Galactic mid-plane. The assumption of a constant force $F_{\Xs}$ at the stream's position is also motivated by the small volume that the stream occupies.

The variation of the potential as a function of height is related to the total matter density of the Galactic disk, written $\rho(\Zs)$, according to Poisson's equation,
\begin{equation}\label{eq:poisson}
    \frac{\partial^2 \Phi_{\Zs}}{\partial {\Zs}^2}
    + \mathcal{R}
    = 4 \pi G \rho({\Zs}).
\end{equation}
The vertical potential component $\Phi_{\Zs}$ is normalised to zero in the mid-plane (where $\Zs=0$), both in function value and first order derivative. The quantity $\mathcal{R}$ is the ``radial term'', and is equal to
\begin{equation}
    \mathcal{R} \equiv 
    \frac{1}{R} \frac{\partial}{\partial R} \left( R \frac{\partial \Phi}{\partial R} \right),
\end{equation}
where $R$ is the Galactic radius. The quantity $R\, \partial \Phi / \partial R$ is equal to the square of the circular velocity, whose derivative with respect to radius is small. For the global Galactic model with which the mock data stellar streams are generated, the radial term can be approximated as a constant, constituting only $\sim 2\%$  of the total mid-plane matter density. It is assumed to be known to good precision, and is corrected for in our model of inference.

\section{Perfect orbit}\label{sec:perf_orbit}
In this section, we first demonstrate the general principles and proof of concept of our method by applying a simplified version of our model of inference to a stellar stream consisting of stars lying on a perfect orbit. This idealised example is useful for demonstrating the merits and limitations of our method, in terms of probing the total matter density distribution of the Galactic disk.

The perfect orbit is generated using the Galactic potential model described in the beginning of Section~\ref{sec:gal_model}. The orbit is perpendicular to the Galactic plane; it has a vertical velocity of $\Ws \simeq -200~\kmsec$ and its velocities parallel to the plane ($\Us$ and $\Vs$) are zero in the Galactic mid-plane. We generate a set of 100 stars, randomly placed at heights smaller than 400 pc from the Galactic mid-plane ($|\Zs|<400~\pc$). 
In this idealised case, we assume to have perfect knowledge of the height of all stars, but we introduce a small uncertainty of $\sigma_{\Ws} = 0.1~\kmsec$ to the vertical velocities. These vertical velocity uncertainties can be interpreted as observational uncertainties, or as an intrinsic velocity dispersion of the stream itself.

In our model of inference, we express the vertical velocity $\Ws$ of stream stars as a function of height above the plane $\Zs$, according to
\begin{equation}\label{eq:w_of_z}
    \Ws^2(\Zs) = \Ws_{0}^2 + 2\Phi_{\Zs}(\Zs),
\end{equation}
where $\Ws_0$ is the vertical velocity in the Galactic mid-plane. For this ideal orbit, we consider a model with four free parameters (encapsulated in $\popp$): the mid-plane velocity ($W_0$), and three parameters that describe the total matter density as a function of height ($\rho_A$, $\rho_B$, $h_A$), according to
\begin{equation}
    \rho(\Zs) =
    \frac{\rho_A}{ \cosh^2 ( \Zs/h_A ) }
    + \rho_B.
\end{equation}
This functional form for the total matter density is quite free to vary in shape, and can also be very close to the true shape of the underlying Galaxy model. Following equation~\eqref{eq:poisson}, this sets the gravitational potential to
\begin{equation}\label{eq:pot_z}
    \Phi_{\Zs}(\Zs,\popp) =
    4 \pi G 
    \rho_A h_A^2 \ln \bigg[ \cosh \bigg( \frac{\Zs}{h_A} \bigg) \bigg]
    + \frac{4 \pi G \rho_B-\mathcal{R}}{2} \Zs^2.
\end{equation}

The posterior density is proportional to
\begin{equation}\label{eq:post_no_errors}
\begin{split}
    & \pr(\popp \, | \, \Zs_{i=\{1,...,100\}},\Ws_{i=\{1,...,100\}}) \propto \\
    & \pr(\popp)\prod_{i=1}^{100}
    \mathcal{N}\left[
    \Ws_{i} \,|\, \text{sign}(\Ws_0)\times\sqrt{\Ws_{0}^2 + 2\Phi_{\Zs}(\Zs_i,\popp)},\, \sigma_{\Ws}
    \right],
\end{split}
\end{equation}
where $\pr(\popp)$ is a flat box prior that takes a non-zero value for $\rho_A,\rho_B \in [0,0.2]~\Msunppcc$, $h_A \in [100,400]~\pc$, and $\Ws_0 \in [-500,500]~\kmsec$. The quantities $\Zs_i$ and $\Ws_i$ are the height and vertical velocity of the $i$th star, and $\mathcal{N}$ is the normal distribution defined as
\begin{equation}\label{eq:normal}
    \mathcal{N}(x \,|\, \bar{x},\sigma) \equiv
    \dfrac{ \exp\left[
    -\dfrac{(x-\bar{x})^2}{2\sigma^2}
    \right]}{\sqrt{2\pi\sigma^2}}.
\end{equation}

\begin{figure}
    \centering
	\includegraphics[width=1.\columnwidth]{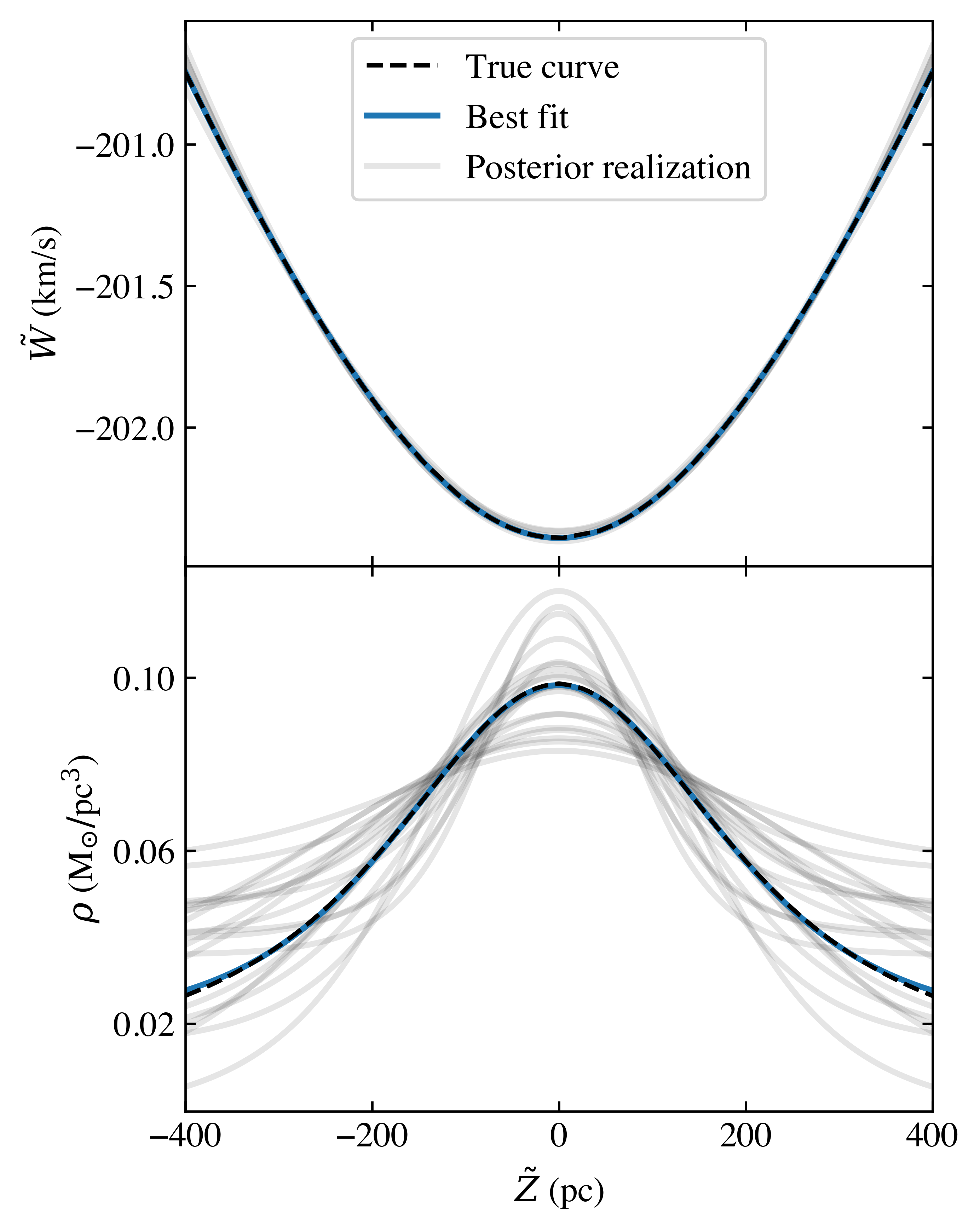}
	\vspace{-0.70cm}
    \caption{The inferred vertical velocity distribution ($\Ws$, upper panel) and total matter density ($\rho$, lower panel), for stars on a perfect orbit. The horizontal axis shows height ($\Zs$) and is shared between both panels. The dashed line represent the true curves for the stars' vertical velocity and the matter density as as a function of height. The solid line represents the best fit curve, found in the limit of zero velocity uncertainties, which is in good agreement with the underlying model. The fainter grey lines are randomly drawn realisations from the posterior density, demonstrating that even with a very small velocity dispersion (in this case 0.1 km/s), information about the shape of the density distribution is poor. The legend applies to both the panels.
    }
    \label{fig:perf_orb_fit}
\end{figure}

\begin{figure}
    \centering
	\includegraphics[width=1.\columnwidth]{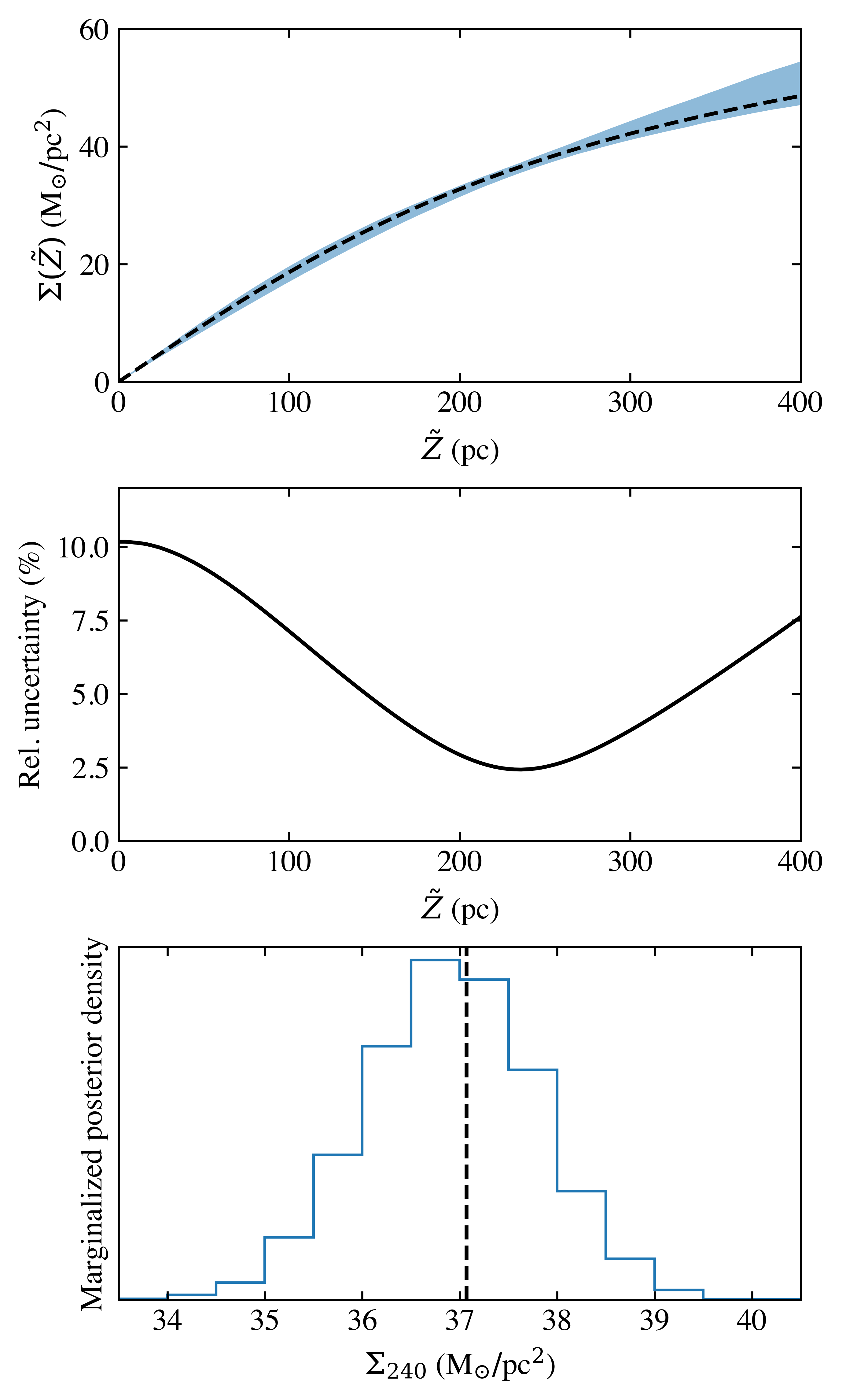}
	\vspace{-0.70cm}
    \caption{Upper panel: Inferred matter surface density within height $\tilde{Z}$, for stars on a perfect orbit, where the band correponds to the 16th and 84th percentiles of the posterior distribution, and the dashed line corresponds to the true value. Middle panel: Relative uncertainty of the inferred $\Sigma(\tilde{Z})$, defined as the posterior standard deviation over its true value. Lower panel: Posterior density for the integrated surface density within 240 pc of the Galactic mid-plane ($\Sigma_{240}$). The dashed line corresponds to the true value. Despite the poor inference on the shape of the density distribution (see Fig.~\ref{fig:perf_orb_fit}), $\Sigma_{240}$ is well constrained to within an uncertainty of $\sim 2.4~\%$.}
    \label{fig:perf_orb_surfdens}
\end{figure}

The inferred matter density distribution is shown in Fig.~\ref{fig:perf_orb_fit}. The fainter grey lines are randomly drawn realisations of the posterior density, which is sampled using a Markov chain Monte Carlo algorithm. Even though the velocity uncertainties are very small ($\sigma_{\Ws} = 0.1~\kmsec$), information about the shape of the density distribution is poor.
We also show the ``best fit'' curve, corresponding to the matter density distribution found in the limit of zero velocity uncertainties ($\sigma_{\Ws}\rightarrow 0~\kmsec$). This curve is in good agreement with the true underlying model, demonstrating that inference on $\rho(\Zs)$ is precise in the fully idealised case, and that the assumed separability and form of the gravitational potential used in the model of inference is valid this close to the Galactic plane.

While the shape of the matter density distribution is poorly constrained, there is a
strong degeneracy between the mid-plane matter density and the distribution's scale height (i.e. between the maximum value of $\rho$ and how quickly $\rho$ decreases with $|\Zs|$). This is similar and analogous to, for example, the inferred mass profile of dwarf galaxies \citep{2010MNRAS.406.1220W,2011ApJ...742...20W}. As seen in Fig.~\ref{fig:perf_orb_fit}, the density is best constrained at a height of around $|\Zs| = 120
~\pc$, corresponding to an uncertainty of around 4~\%. However, a more precisely determined quantity is the integrated surface density within 240 pc of the mid-plane. The surface density, integrated to a height of $\tilde{Z}$, is defined like
\begin{equation}
    \label{eq:surface_density}
    \Sigma(\tilde{Z}) \equiv \int_{-\tilde{Z}}^{\tilde{Z}} \rho(\Zs') \de \Zs',
\end{equation}
and is shown in the top panel of Fig.~\ref{fig:perf_orb_surfdens}. The middle panel shows the relative uncertainty of the inferred $\Sigma(\tilde{Z})$, which has a minimum at $\tilde{Z} = 240~\pc$.

In the lower panel of Fig.~\ref{fig:perf_orb_surfdens}, we show the inferred surface density within 240 pc from the Galactic mid-plane, written $\Sigma_{240}\equiv \Sigma(240~\pc)$ for shorthand. The marginalised posterior has a mean and standard deviation of $\Sigma_{240} = (36.91 \pm 0.88)~\Msunppcsquare$, corresponding to an uncertainty of 2.4 \% with respect to its true value of $37.07~\Msunppcsquare$. In summary, it is difficult to infer the shape of matter density distribution, but the integrated surface density within $240~\pc$ can still be inferred with good precision.

\section{Stream model}\label{sec:model}

In this section, we describe the model of inference used in the rest of this work. Unlike the idealised case presented in the previous section, here we will consider realistic stellar streams that possess an intrinsic phase-space dispersion, as well as realistic observational uncertainties and incomplete radial velocity information. In order to model such a stream, it becomes necessary to describe its full six-dimensional phase-space density, rather than just its vertical components.

As demonstrated in Section \ref{sec:perf_orbit}, it is difficult to constrain the shape of matter density distribution, even for streams with small intrinsic velocity dispersion. For this reason, we fix the shape of the total matter distribution in our model of inference, and parametrize it in terms of the surface density within $240~\pc$ from the Galactic plane ($\Sigma_{240}$), because this quantity could be most accurately measured. For the rest of this article, the matter distribution depends only on one free parameter, rather than three free parameters as in Section~\ref{sec:perf_orbit}. The matter density as a function of height, in units $\Msunppcc$, is equal to
\begin{equation}
    \rho(\Zs) =
    \frac{\Sigma_{240}}{382.77~\pc}
    \left[
    \frac{0.8}{ \cosh^2 ( \Zs/h_A ) }
    + 0.2
    \right],
\end{equation}
where $h_A = 230~\pc$. The value $382.77~\pc$ is a normalisation that is given by the fixed shape of the matter density distribution. This function is close to the true shape of the matter distribution of the Galactic model (visible in the lower panel of Fig.~\ref{fig:perf_orb_fit}).

{\renewcommand{\arraystretch}{1.6}
\begin{table*}
	\centering
	\caption{Model parameters and data of an observed star. The 16 model parameters are split into two subsets, $\popp=\{\popp_S,\popp_\odot\}$: the former parametrizes the stellar stream's intrinsic phase-space distribution; the latter parametrizes the stream's position and velocity relative to the Sun.}
	\label{tab:parameters}
    \begin{tabular}{| l | l |}
		\hline
		$\popp_S$  & Model parameter subset describing the stream's phase-space density \\
		\hline
		$\Sigma_{240}$ & Total matter surface density within 240 pc of the Galactic mid-plane \\
		$F_{\Xs}$ & Force per mass acting on the stream in the direction of the Galactic centre \\
		$\Us_0,\, \Vs_0, \,\Ws_0$ & Velocity of the stream orbit's intersection with the Galactic plane \\
		$\sigma_{\{\Us,\Vs,\Ws\}}$ & Three intrinsic velocity dispersions \\
		$\sigma_{\{\Xs,\Ys\}}$ & Intrinsic dispersion in $\Xs$ and $\Ys$, respectively \\
		\hline
		\hline
		$\popp_\odot$  & Model parameter subset describing the Sun's relative position and velocity \\
		\hline
		$d_0$ & Distance to the stream from the Sun \\
		$l_0$ & Galactic longitude of the stream at its intersection with the Galactic plane \\
		$Z_\odot$ & Height of the Sun with respect to the Galactic mid-plane \\
		$U_\odot,\, V_\odot,\, W_\odot$ & Velocity of the Sun with respect to the Galaxy \\
        \hline
        \hline
        $\data_i$ & Data of a star with index $i$ \\
        \hline
        $\hat{l}_i,\, \hat{b}_i$ & Galactic longitude and latitude (with negligible uncertainties) \\
        $\hat{\varpi}_i,\, \hat{\sigma}_{\varpi,i}$ & Observed parallax and associated uncertainty \\
        $\hat{\mu}_{l,i},\, \hat{\mu}_{b,i},\, \hat{\sigma}_{\mu,i}$  & Observed proper motions and associated uncertainty \\
        $\hat{v}_{\text{RV},i},\, \hat{\sigma}_{\text{RV},i}$ & Observed radial velocity and associated uncertainty (potentially available) \\
        \hline
	\end{tabular}
\end{table*}}

Our model contains 16 free parameters, which are also listed in Table~\ref{tab:parameters}: the surface density ($\Sigma_{240}$), as mentioned above; the gravitational force per mass ($F_{\Xs}$) acting on the stream in the direction of the Galactic centre, see equation~\eqref{eq:potential}; the three-dimensional velocity of the stream as it passes through the Galactic mid-plane ($\Us_0$, $\Vs_0$, $\Ws_0$); five parameters that describe the stream's intrinsic phase-space dispersion ($\sigma_{\{\Us,\Vs,\Ws\}}$, $\sigma_{\{\Xs,\Ys\}}$); six parameters that describe the Sun's position and velocity relative to the stream ($d_0$, $l_0$, $Z_\odot$, $U_\odot$, $V_\odot$, $W_\odot$). The full set of free parameters are split into two sets: the first ten parametrize the stellar stream's intrinsic phase-space distribution ($\popp_S$); the latter six parametrize the stream's position and velocity with respect to the Sun ($\popp_\odot$).

The orbit of the stellar stream is parametrized by its height with respect to the plane ($\Zs$); all other phase-space coordinates ($\Xs$, $\Ys$, $\Us$, $\Vs$, $\Ws$) are functions of $\Zs$, using the ``stream frame'' coordinate system defined in Section~\ref{sec:gal_model}. Just like in Section~\ref{sec:perf_orbit}, the gravitational potential of the model of inference is separable according to equation~\eqref{eq:potential}, and the vertical velocity is described by equation~\eqref{eq:w_of_z}. There is no gravitational force acting in the azimuthal direction. There is a constant force per mass in the direction of positive $\Xs$, such that the velocity $\Us$ changes linearly with time. Thus the velocities of the stream's orbit follow
\begin{equation}\label{eq:vels_model}
    \begin{split}
        \Us(\Zs) &= \Us_0 + F_{\Xs} \tilde{t}(\Zs),\\
        \Vs(\Zs) &= \Vs_0,\\
        \Ws^2(\Zs) &= \Ws_{0}^2 + 2\Phi_{\Zs}(\Zs)
    \end{split}
\end{equation}
where
\begin{equation}
    \tilde{t}(\Zs) = \int_{0}^{\Zs} \frac{\de \Zs'}{\Ws(\Zs')}.
\end{equation}
is the time that has passed between a stars current position and its mid-plane passage (which is negative if the star is approaching the mid-plane). Integrating this with respect to time gives the spatial positions
\begin{equation}
    \begin{split}
        \Xs(\Zs) &= \Xs_0 + \Us_0 \tilde{t}(\Zs) + \frac{1}{2} F_{\Xs} \tilde{t}^2(\Zs),\\
        \Ys(\Zs) &= \Ys_0 + \Vs_0 \tilde{t}(\Zs),
    \end{split}
\end{equation}
where $\Xs_0$ and $\Ys_0$ are the positions where the orbit of the stream passes the mid-plane (given by the model parameters $\popp_\odot$, see Appendix~\ref{app:transformations} for details).

The stream stars have an intrinsic scatter in phase-space around this ideal orbit, assumed to be Gaussian and modelled by dispersions $\sigma_{\{\Us,\Vs,\Ws,\Xs,\Ys\}}$. The full phase-space density for a stream star, denoted with an index $i$, is thus proportional to
\begin{equation}\label{eq:phase_space_density}
\begin{split}
& f(\boldsymbol{\Xs_i}, \boldsymbol{\Vs}_i \, | \, \popp_S) \propto \mathcal{N}[\Us_i \,|\, \Us(\Zs_i),\, \sigma_{\Us}] \\
& \times \mathcal{N}(\Vs_i \,|\, \Vs_0,\, \sigma_{\Vs}) \times \mathcal{N}[\Ws_i \,|\, \Ws(\Zs_i),\, \sigma_{\Ws}] \\
& \times \mathcal{N}[\Xs_i \,|\, \Xs(\Zs_i),\, \sigma_{\Xs}] \times \mathcal{N}[\Ys_i \,|\, \Ys(\Zs_i),\, \sigma_{\Ys}].
\end{split}
\end{equation}

Given a set of $N$ stream stars, labelled by the index $i$, the posterior density of the model is written
\begin{equation}\label{eq:posterior}
\begin{split}
    & \pr(\popp \, | \, \data_{i=\{1,...,N\}}) \propto \\
    & \pr(\popp)\prod_{i=1}^N \int \pr(\data_i \, | \, \boldsymbol{\Xs}_i,\, \boldsymbol{\Vs}_i,\, \popp_\odot) \\
    & \times f(\boldsymbol{\Xs}_i,\boldsymbol{\Vs}_i \, | \, \popp_S)\, \de^3 \boldsymbol{\Xs}_i \,\de^3 \boldsymbol{\Vs}_i,
\end{split}
\end{equation}
where $\pr(\popp)$ is a prior on the 16 model parameters, $\pr(\data_i \, | \, \boldsymbol{\Xs}_i,\, \boldsymbol{\Vs}_i,\, \popp_\odot)$ is the likelihood of the data of the $i$th star, and $f(\boldsymbol{\Xs}_i,\boldsymbol{\Vs}_i \, | \, \popp_S)$ is the phase-space density of the stream, as expressed in equation~\ref{eq:phase_space_density}. The posterior contains $N$ six-dimensional phase-space integrals, and also depends on quite a high number of free parameters, which is described in detail in Appendix~\ref{app:posterior}. However, the posterior can be significantly simplified by reducing the six-dimensional integrals analytically, to numerical integration in only one dimension (see Appendix~\ref{app:reduction}). This posterior is then implemented in \texttt{TensorFlow} \citep{tensorflow2015-whitepaper}, which enables auto-differentiation of the posterior with respect to its free parameters. This allows for efficient minimisation as well as Hamiltonian Monte-Carlo (HMC) sampling of the posterior density, despite its high dimensionality. The sampling strategy is described in greater detail in Appendix~\ref{app:sampling}. All the code that is used to produce the results in this paper is open source and can be found online.\footnote{\url{https://github.com/AxelWidmark/wool}}

The parameter of interest is the surface density parameter $\Sigma_{240}$, and all other parameters of the model are largely to be regarded as nuisance parameters. The parameters that describe the Sun's position and velocity ($Z_\odot$, $U_\odot$, $V_\odot$, $W_\odot$) and the gravitational force acting on the stream in the direction of the Galactic centre ($F_{\Xs}$) are more precisely and accurately determined by other methods. However, they are still included as free parameters, as the uncertainty with respect to these parameters can inflate the uncertainty of $\Sigma_{240}$. Knowledge about these parameters is reflected by the prior, which is described in more detail in Appendix~\ref{app:prior}.

\section{Mock data stellar streams}\label{sec:sim_streams}

Stellar streams are formed by the tidal disruption and dissolution of satellites, where the escaping stars are lifted out of the gravitational potential well of their progenitor. In doing so, the stars end up with slightly different energies and momenta, causing an intrinsic phase-space dispersion of the stream. In addition to this, the measurements of phase-space positions of stars is accompanied by observational uncertainties, and sometimes missing information (in this case incomplete radial velocity measurements). In this section, we describe how we generate our mock data stellar streams, employing realistic phase-space dispersions and observational uncertainties.

We generate four mock data stellar streams containing 300 stars each, using the Galactic potential described in the beginning of Section~\ref{sec:gal_model}. The streams are generated as stars on a perfect orbit, randomly placed along its spatial trajectory. The stars are then displaced in phase-space, using dispersions of 20 pc in all three spatial directions, and 1 km/s for all three velocities. These values are adopted in accordance with the internal phase-space dispersions of some of the known dynamically cold streams of the Milky Way \citep{GrillmairCarlin2016}, and also match the values obtained in some recent studies \citep{2019MNRAS.486.2995M}. We neglect any correlations between the spatial and velocity offsets with respect to the ideal orbit. Such correlations, if they exist, could correspond to a conserved total energy or angular momentum (which would be of order 0.1 km/s for the vertical velocity) or a lower velocity dispersion in the outskirts of the stellar stream (a correction which would be of similar magnitude). These effects are expected to be negligible and sub-dominant to the observational uncertainties of the vertical velocity, which is discussed in more detail towards the end of this section. Another potential deviation with respect to the ideal orbit is energy segregation between the leading and trailing arms of the stellar stream. This segregation is typically small for a low-mass progenitor, especially if the progenitor is completely dissolved. It can cause a significant bias to the inferred matter density if the segregation is large with respect to the intrinsic velocity dispersion of the stellar stream. Accounting for such effects is discussed in the end of Section~\ref{sec:results}.

In order to be able to make simple comparisons of the inferred surface density using the different mock data stellar streams, they are all located at the same Galactic radius of 8.1 kpc, where the rotational velocity is roughly 232 km/s (thus we move the position from which the stream is observed, rather than the stream itself). The vertical velocity of all streams is negative, such that their direction of motion is towards the Galactic south. All streams are observed at a distance of about 1 kpc. The four streams are named S1--S4, and their properties are described below.

\textbf{Stream S1} has velocities $\boldsymbol{\Vs} \simeq \{0,\, 0,\, -100\}~\kmsec$. It is somewhat idealised, in the sense that it passes perpendicularly through the Galactic disk, with a rather small vertical velocity, such that the stream is at the apocentre of an eccentric orbit. Its total extent is $\sim 800~\pc$, with heights between $\Zs \in [-400,400]~\pc$. In the observer's frame, the median position of its stars is located at $\{X,Y\} = \{0,1\}~\kpc$.

\textbf{Stream S2} is on a prograde orbit, with velocities $\boldsymbol{\Vs} \simeq \{15,\, 220,\, -45\}~\kmsec$. Its total extent is $\sim 1~\kpc$, with heights between $\Zs \in [-450,-200]~\pc$. In the observer's frame, the median position of its stars is located at $\{X,Y\} = \{-1,0\}~\kpc$.

\textbf{Stream S3} is also on a prograde orbit, with velocities $\boldsymbol{\Vs} \simeq \{0,\, 200,\, -70\}~\kmsec$. Its total extent is $\sim 500~\pc$, with heights between $\Zs \in [200,400]~\pc$. In the observer's frame, the median position of its stars is located at $\{X,Y\} = \{-1,0\}~\kpc$.

\textbf{Stream S4} has velocities $\boldsymbol{\Vs} \simeq \{-160,\, 160,\, -70\}~\kmsec$. Its orbit is inclined with respect to both the Galactic plane and the direction of rotation. Its total extent is $\sim 500~\pc$, with heights between $\Zs \in [200,400]~\pc$. In the observer's frame, the median position of its stars is located at $\{X,Y\} = \{0,1\}~\kpc$.

The intrinsic phase-space coordinates of these streams are shown in Fig. \ref{fig:coords_S1}--\ref{fig:coords_S4}, found in Appendix~\ref{app:coords}.

The intrinsic phase-space coordinates of the streams are transformed into observables, given by the relative phase-space position of the Sun. An individual stream star, denoted by the index $i$, has the following observables: angular position on the sky ($\hat{l}_i$, $\hat{b}_i$), parallax ($\hat{\varpi}_i$), proper motions ($\hat{\mu}_{l,i}$, $\hat{\mu}_{b,i}$), and (for a subset of sufficiently bright stars) radial velocity ($\hat{v}_{\text{RV},i}$). They all have their associated uncertainties, with the exception of the angular position on the sky, whose uncertainties are negligible. This constitutes the data ($\data_i$) of an observed star, which is listed in Table~\ref{tab:parameters}.

\begin{figure}
    \centering
	\includegraphics[width=1.\columnwidth]{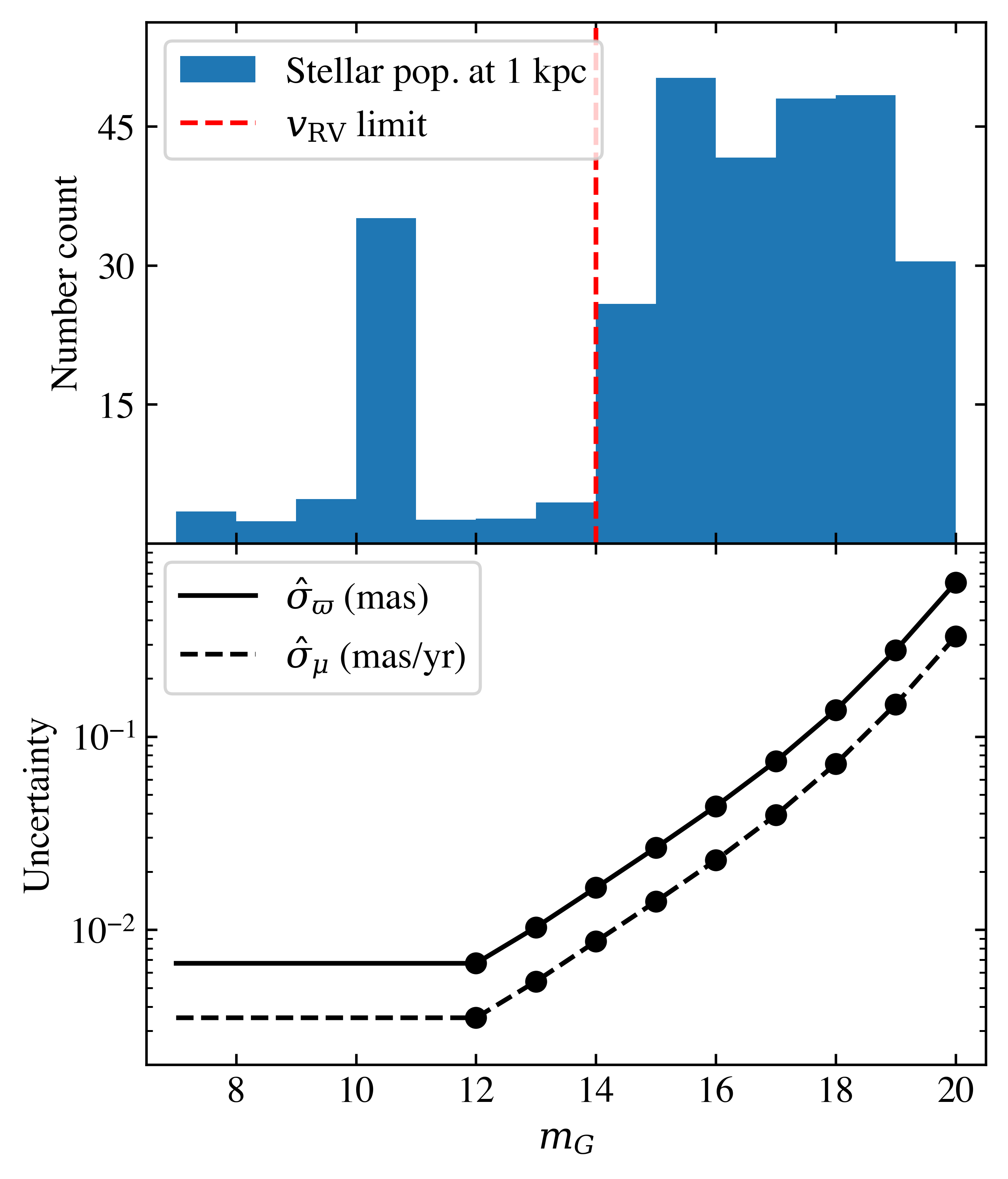}
	\vspace{-0.70cm}
    \caption{Expected apparent magnitudes for stars at a distance of 1 kpc (top panel), and associated \emph{Gaia} end-of-mission parallax and proper motion uncertainties (bottom panel, units for the vertical axis are written in the legend). The sample of stars is obtained from a synthesised globular cluster population with iron abundance $[\text{Fe/H}] = -2.0$ and an age of 12 Gyr, and the histogram is normalised to a sum total of 300 stars. The upper panel also shows the apparent magnitude limit for which radial velocities ($v_\text{RV}$) are available (18.5 \% of stars in this population). The solid circles in the bottom panel correspond to the interpolation points of the \emph{Gaia} performance forecast. The horizontal axis is shared between the two panels.}
    \label{fig:errors}
\end{figure}

We use the expected \emph{Gaia} end-of-mission uncertainties, as a function of the apparent $G$-band magnitude $m_G$. In order to simulate a realistic distribution of apparent magnitudes, we synthesise a typical globular cluster population with an iron abundance of $[\text{Fe/H}] = -2.0$ and an age of 12 Gyr, using \textsc{Isochrone} \citep{isochrones}. When generating our mock data stellar streams, each star is randomly assigned an absolute magnitude, from which the apparent magnitude is given by its distance modulus. The synthesised distribution of apparent magnitudes, for stars at a distance of 1 kpc, is visible in the top panel of Fig.~\ref{fig:errors}, where the peak in the histogram around $m_G=10.5$ corresponds to blue horizontal branch stars. The histogram is normalised to a sum total of 300 stars (although it was generated using 4000 stars, in order to reduce the statistical noise of bin heights). Given this distribution of magnitudes, the mock data stellar streams have an approximate surface brightness of $\sim 28 \text{mag}/\arcs^2$; this is consistent with many known stellar streams of the Milky Way \citep{2008ApJ...689..936J,2016ASSL..420..219C}, although recent surveys (e.g. \emph{Gaia}, \citealt{2016A&A...595A...1G}, DES, \citealt{2018PhRvD..98d3526A}) have discovered plenty of fainter streams (30--34 $\text{mag}/\arcs^2$) in our Galaxy \citep{2018ApJ...862..114S,2018ApJ...865...85I}.

The expected parallax and proper motion uncertainties ($\hat{\sigma}_{\varpi}$ and $\hat{\sigma}_{\mu}$) are shown in the bottom panel of Fig.~\ref{fig:errors}.\footnote{The interpolation points are taken from this table: \url{https://www.cosmos.esa.int/web/gaia/sp-table1}} For the spectrograph that measures radial velocity, the expected limit in apparent magnitude is $m_G < 14.5$ \citep{2016A&A...585A..93R}, although this is a function of position on the sky. We are interested in streams passing through the Galactic disk, where stellar crowding can impede the spectrograph's performance. For this reason we set a slightly more conservative limit of $m_G < 14$. We set the radial velocity uncertainty to $\hat{\sigma}_{\text{RV}}= 0.3$ km/s \citep{2019A&A...622A.205K}. The number of stars with available radial velocities for S1--S4 are 60, 61, 58, and 64, respectively. Stellar crowding can also affect the precision of other astrometric measurements, as well as stellar completeness. We do not expect this to cause a significant effect, because stars close to the Galactic mid-plane are less important for inferring the surface density; what is crucial is measuring the slope of $\Ws$ with respect to $\Zs$, which is steeper at greater values of $|\Zs|$. Furthermore, most of the mock data stellar streams (with the exception of S1) do not actually pass through Galactic mid-plane; for such streams, the astrometric precision and stellar completeness should not be significantly impeded.

The main limiting factors for measuring the matter density of the Galactic disk are the parallax uncertainties and the stream's intrinsic phase-space dispersion. The parallax uncertainties of the mock data stellar streams have a median value of $\hat{\sigma}_\varpi \simeq 0.05~\mas$. This corresponds to a 5 \% uncertainty at a distance of 1 kpc, which is larger than the stream's spatial dispersion of 20 pc. However, the spatial extent of the stellar stream is well constrained by the subset of bright stars for which the measurements are more precise, which in turn constrains the position of stars for which parallax information is poor. This is an effect of Bayesian deconvolution (similar in principle to for example \citealt{2018AJ....156..145A}).

For most proper motion measurements, the observational uncertainty is largely negligible. However, because uncertainty in distance is large (at best roughly $10~\pc$, and for most stars at least $20~\pc$), this translates into an uncertainty in a star's vertical velocity. Given a vertical velocity of about $\Ws = 70~\kmsec$, the uncertainty for a single star's vertical velocity is typically
\begin{equation}
    \Delta \Ws \gtrsim 70~\kmsec \times \left( \frac{20~\pc}{1~\kpc} \right) = 1.4~\kmsec.
\end{equation}
In summary, the vertical velocity of a star is at best constrained to about 1 km/s. In addition to this observational difficulty, the streams also have an intrinsic velocity dispersion of 1 km/s. For this reason, the phase-space correlations discussed in the beginning of this section, which are of the order $0.1~\kmsec$, are sub-dominant and have a negligible effect.

\section{Results}\label{sec:results}

We explore the posterior density of our four mock data stellar streams using Hamiltonian Monte-Carlo (HMC) sampling, described in Appendix \ref{app:sampling}.

\begin{figure}
    \centering
	\includegraphics[width=1.\columnwidth]{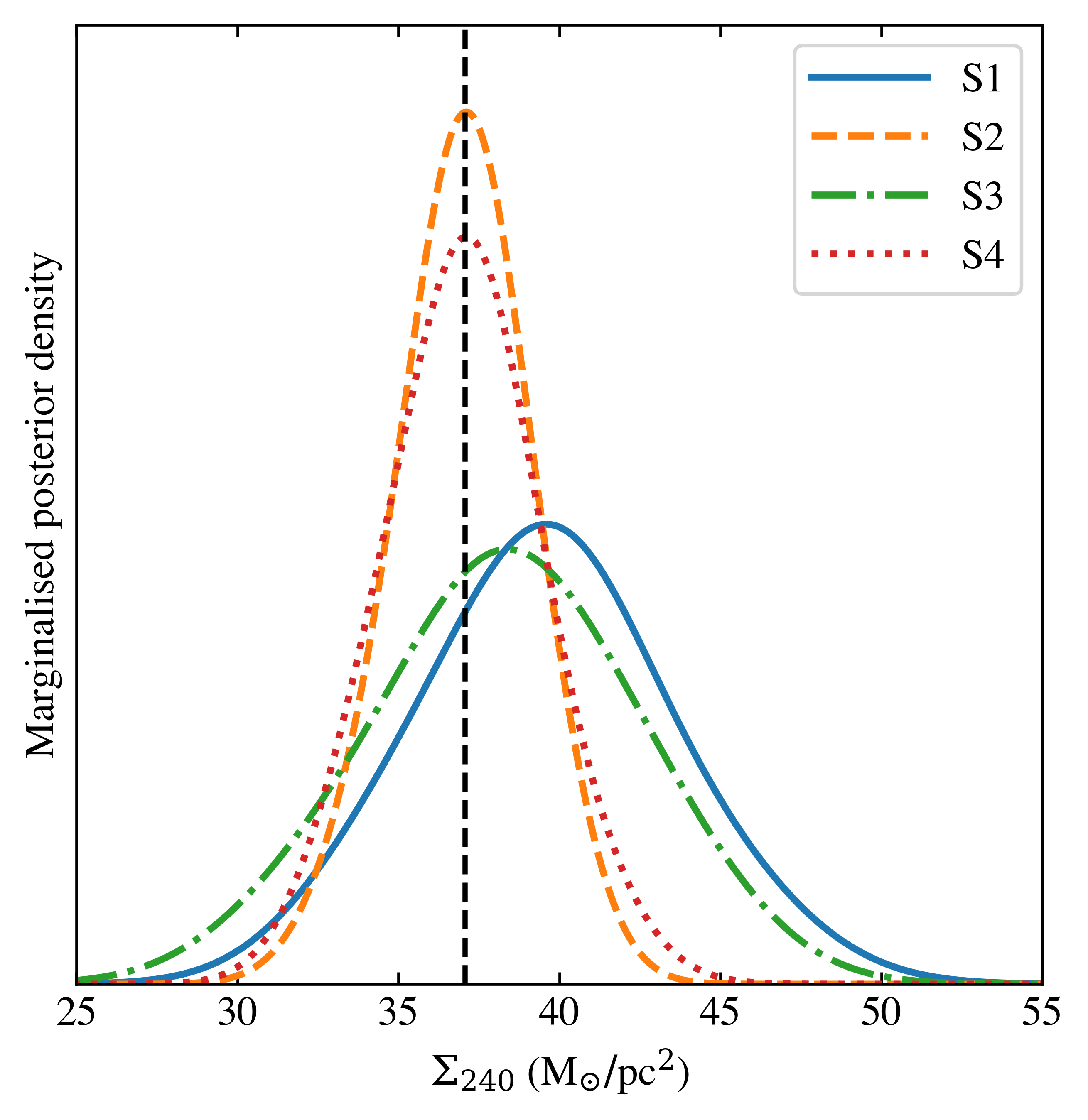}
	\vspace{-0.70cm}
    \caption{Marginalised posterior density for the integrated surface density within 240 pc of the mid-plane ($\Sigma_{240}$), for streams S1--S4. The dashed vertical line corresponds to the true surface density of the Galactic model used to generate the mock data stellar streams.}
    \label{fig:posterior}
\end{figure}

The inferred posterior for the integrated surface density within 240 pc of the mid-plane ($\Sigma_{240}$), for all four streams, is shown in Fig.~\ref{fig:posterior}. The orbits of these streams are quite varied, but the inference of the surface density is accurate for all streams. The posterior width of the inferred surface density varies from roughly 6--12 \% of the true value. Other mock data realisations of the same stellar streams produce similar results, with small shifts to the median value of $\Sigma_{240}$.

\begin{figure*}
    \centering
	\includegraphics[width=1.\textwidth]{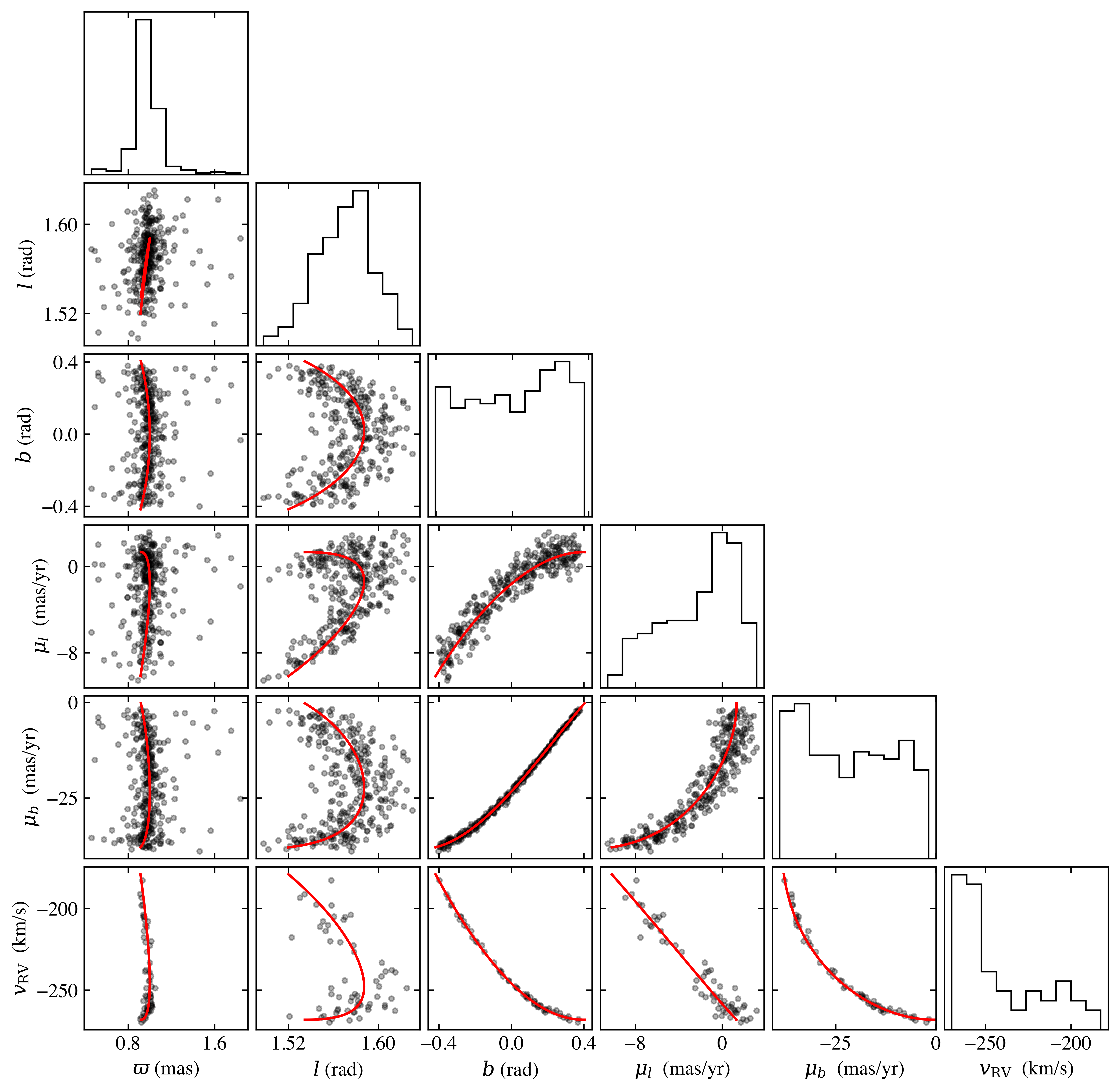}
	\vspace{-0.70cm}
    \caption{Data and model fit of stream S1. The dots correspond to the observed values of the respective observables, while the solid line is the inferred orbit of the stream, obtained using the median values for all parameters of the posterior density. The radial velocity ($v_\text{RV}$) is plotted only for the subset of stars for which it is available (60 out of 300).}
    \label{fig:data_and_orbit}
\end{figure*}

In Fig.~\ref{fig:data_and_orbit}, we plot the data and model fit for stream S1. The model fit in this figure corresponds to the respective median values of all 16 model parameters of the inferred posterior. The inferred model parameters of the other mock data stellar streams produce similarly good fits with respect to the data.

\begin{figure*}
\centering
\subfloat[S1]{\includegraphics[width=0.45\textwidth]{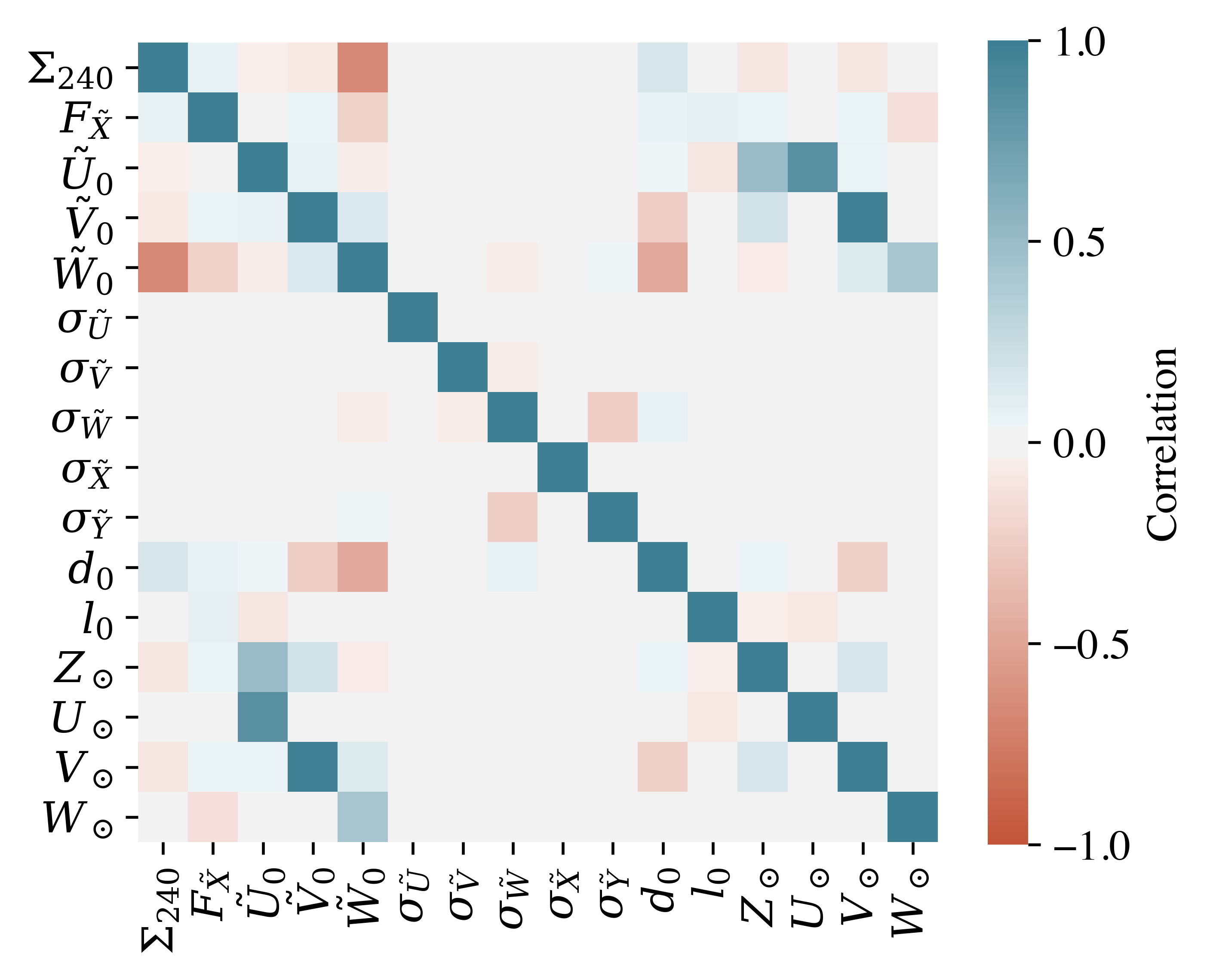}} \hspace{3mm}
\subfloat[S2]{\includegraphics[width=0.45\textwidth]{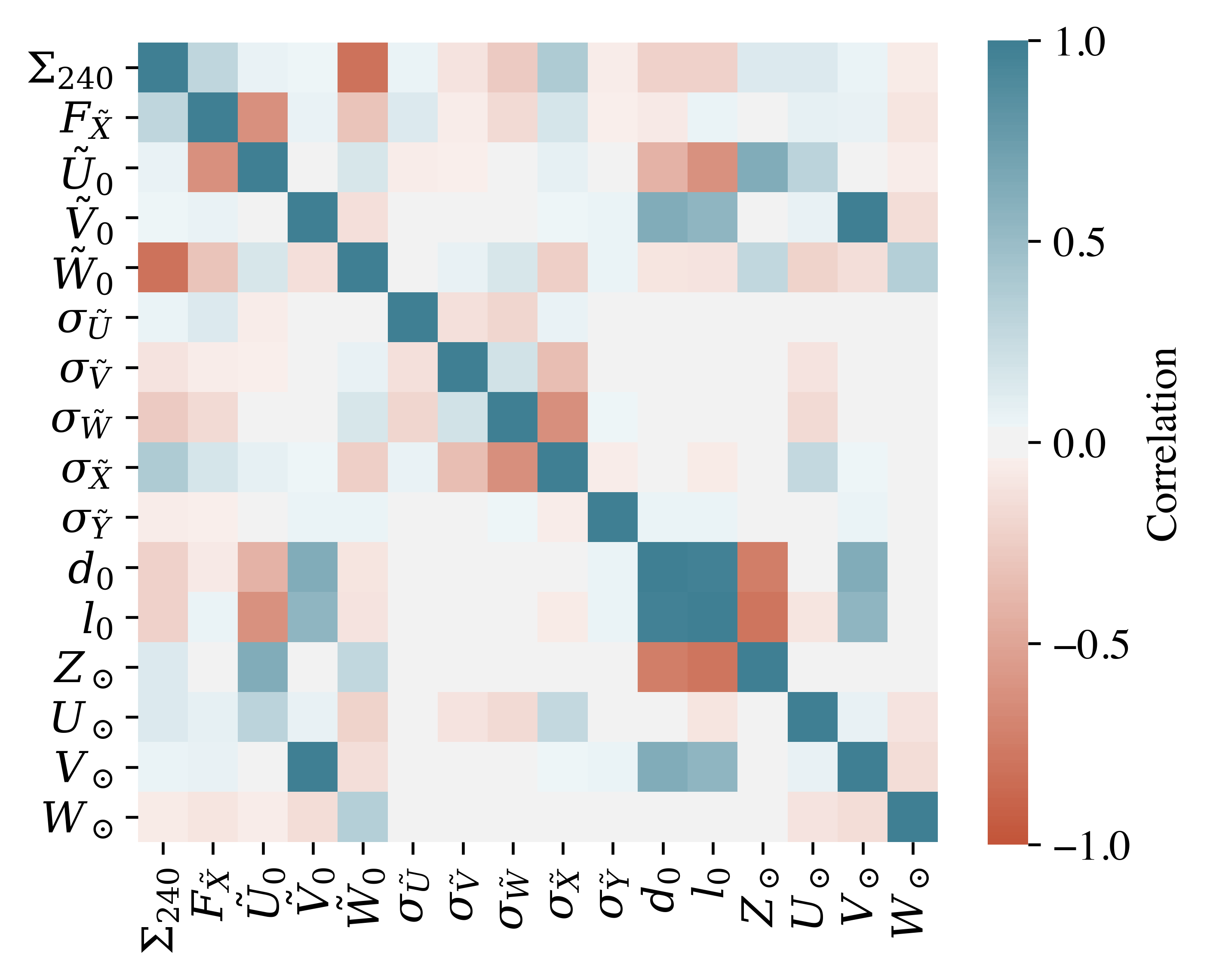}}
\\
\noindent 
\subfloat[S3]{\includegraphics[width=0.45\textwidth]{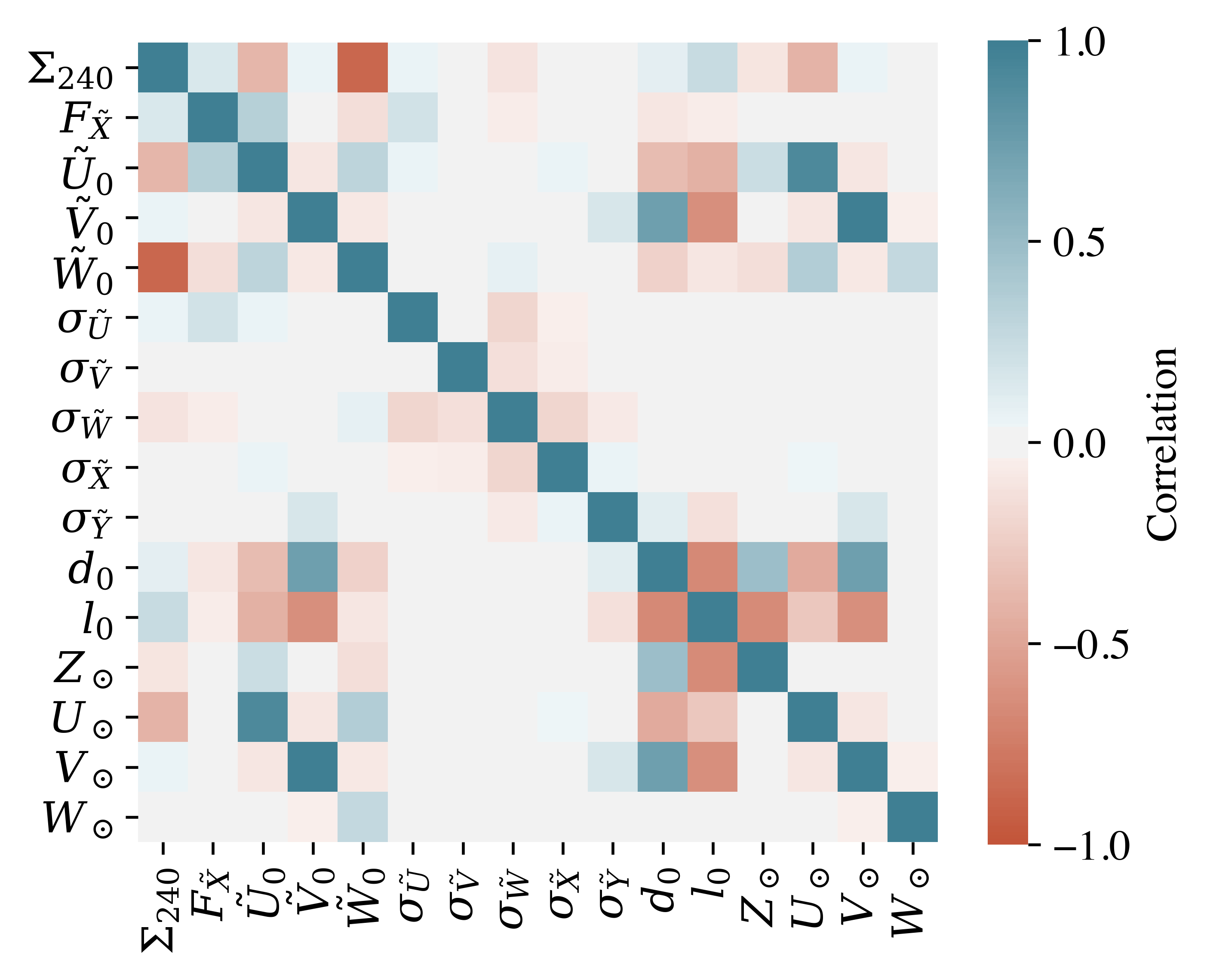}} \hspace{3mm}
\subfloat[S4]{\includegraphics[width=0.45\textwidth]{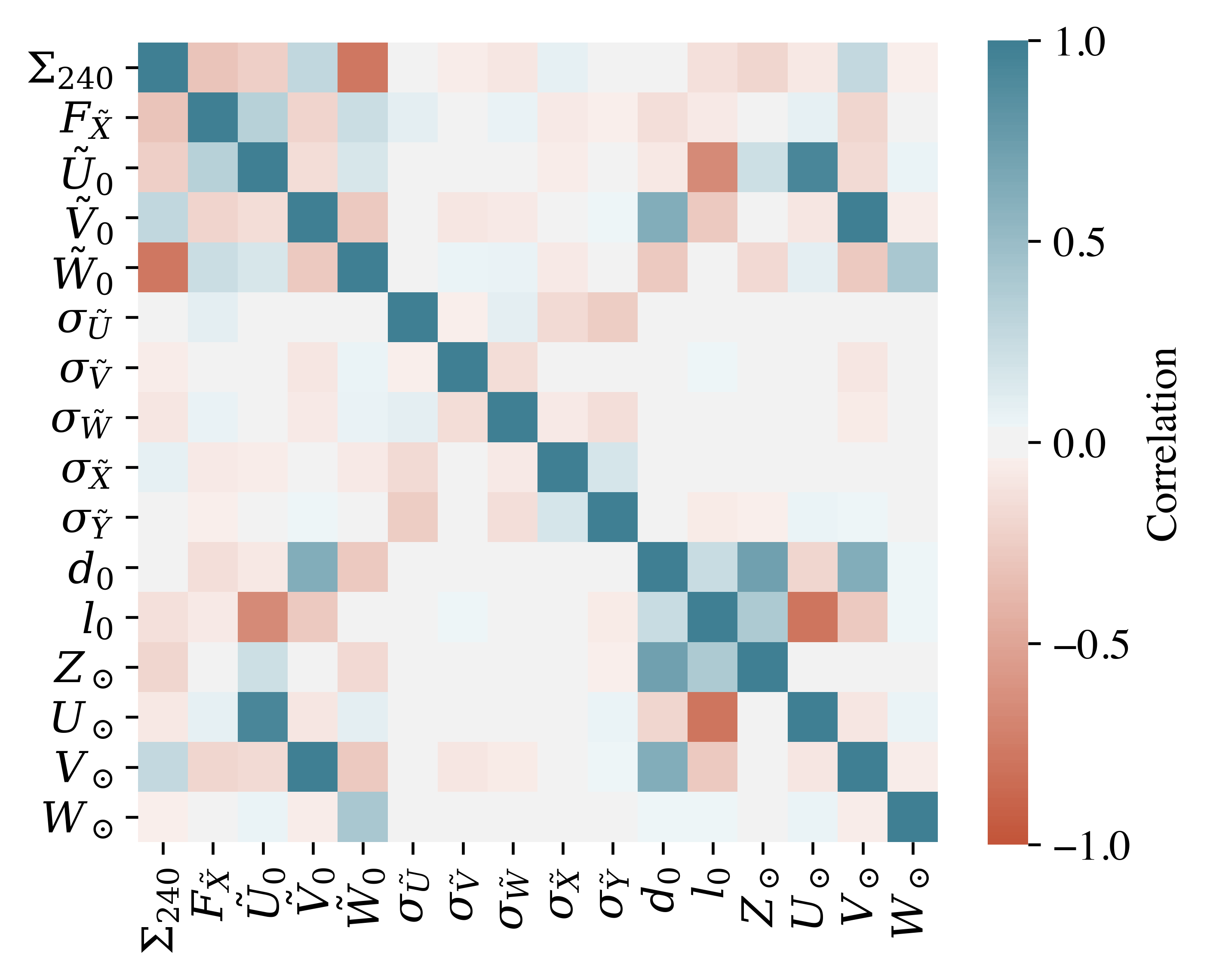}}
\caption{Correlation matrices for the inferred posterior of the four mock data stellar streams (S1, S2, S3, S4).}
\label{fig:correlations}
\end{figure*}

In Fig.~\ref{fig:correlations}, we show a correlation matrix for the model's 16 free parameters, as inferred for streams S1--S4. The parameter of interest, the surface density $\Sigma_{240}$, is not strongly correlated with other parameters, with the exception of some anti-correlation with the stream's vertical velocity $\Ws_0$ (the vertical velocity is negative, so a greater $\Sigma_{240}$ correlates with a greater absolute value of $\Ws_0$). Other strong degeneracies exist between $\Us_0$ and $U_\odot$, and between $\Vs_0$ and $V_\odot$, which parametrizes the solar and stream velocities parallel to the Galactic plane. If there was a strong correlation between $\Sigma_{240}$ and model parameters for which we have a highly constrained prior ($F_{\Xs}$, $Z_\odot$, $U_\odot$, $V_\odot$, $W_\odot$, see Appendix~\ref{app:prior} for more details), then this would indicate high sensitivity to systematics associated with these parameters. In such a case, a prior density that is offset from its true value would translate to a bias in the inferred value for $\Sigma_{240}$. Fortunately, such strong degeneracies are not present in the posterior densities of samples S1--S4.

The precision of inference in terms of the matter density is highly contingent on the stream's orbit. The four mock streams considered in this work have different vertical velocities, ranging from roughly 45 to 100 km/s. With greater vertical velocities the relative change to the vertical energy is smaller ($\delta \Ws \propto \Ws^{-1}$), resulting in lower precision. On the other hand, very low vertical velocities means that the stream does not vary much in $\Zs$. In order to have good precision of inference, a stream with such low inclination would have to be very long, and in such a case the disk density that it is probing is no longer very local. Furthermore, such a stream would be more sensitive to perturbations from structures in the Galactic disk, given its low vertical energy.

The inference discussed above is for an observational distance of roughly 1 kpc, but this method is applicable also for streams at larger distances, as long as the distance to the stream can be accurately constrained. Given the same intrinsic phase-space dispersion, the velocity dispersion is dominant over the proper motion uncertainty out to a distance of several kilo-parsec. We have used a realistic phase-space dispersion, similar to what have been observed for some streams in the Milky Way. There is also the possibility of finding dynamically colder streams, which would improve the precision.

When analysing an actual stream, it is important to examine how well the data agrees with the inferred model, especially in terms of systematic offsets in phase-space with respect to the stream's ideal orbit. Such offsets could arise from an energy segregation of the stream's leading and trailing arms. This is not currently included in the model, and can bias the result in the event that such offsets are significant with respect to the intrinsic phase-space dispersion of the stream. Identifying and quantifying such a feature can be more or less easy to do, depending on the viewing angle of the stream and if the progenitor globular cluster has been disrupted.

\section{Conclusion}\label{sec:discussion}

In this work, we have demonstrated that the total matter density of the Galactic disk can be constrained by analysing the kinematics of a stellar stream that passes through or close to the Galactic mid-plane. This is possible because the vertical energy of stream stars is approximately constant, such that the stars' heights and vertical velocities are interrelated via the gravitational potential. The method was demonstrated on realistic mock data stellar streams, with projected end-of-mission \emph{Gaia} uncertainties. We found that a small stream 
containing 300 stars with realistic phase-space dispersion, situated at a
distance of approximately $1~\kpc$, can be used to constrain the Galactic disk surface density to within an uncertainty of $6~\%$. This level of precision is competitive with other methods using disk tracer stars. Our method accounts for the full error covariance matrix of all individual stars, as well as for the uncertainties associated with the Galactic potential and solar velocities. 
Our method is highly complementary to the methods using disk tracer stars, because it does not rely on any steady-state assumption for any stellar population of the disk itself. Furthermore, it does not rely on stellar number density information, and is therefore insensitive to stellar number count biases arising from, for example, a mismodelled selection function. Our method also makes it possible to make precise measurements of the surface density at different locations in the Galactic disk, wherever a suitable stream can be found. This could give us insights with regards to the disk's dynamical evolution, for instance if the disk is perturbed by the passage of the Sagittarius dwarf galaxy \citep{Laporte2019} or due to buckling of the stellar bar \citep{Khoperskov2019}. Such perturbations could cause spatial variations in the matter density of the Galactic disk, and potentially break axisymmetry in a way that increases the importance of the radial and azimuthal terms of Poisson's equation.

Our method is best suited for low-mass dynamically cold stellar streams, much like those produced from the tidal disruption of globular clusters or very low-mass dwarf galaxies. Streams produced by more massive dwarf galaxies are more diffuse in phase-space, resulting in higher uncertainties in the inferred surface matter density. \emph{Gaia} DR2 has unveiled a large amount of stellar streams in the solar neighbourhood \citep{2018ApJ...860L..11K, 2018MNRAS.475.1537M, 2019A&A...631L...9K, 2019arXiv191007538Y, 2020MNRAS.492.1370B}, although most of these streams have rather large phase-space dispersions, as is expected from tidal debris of more massive progenitors. An exception is \citet{2019A&A...622L..13M}, who found a small stream only $100\pc$ from the Sun. The stream extends at least 400 pc in distance, but is not useful for our purposes, as it has a very low vertical velocity ($\sim 11~\kmsec$) and varies only minimally in height with respect to the Galactic plane.

Most stream finding algorithms mask the region surrounding the disk, in order to bypass the computational challenge of processing a very large number of stars \citep{2018MNRAS.481.3442M,2020MNRAS.492.1370B}. However, several globular clusters are known to orbit the Milky Way close to the disk \citep{2019arXiv190605864A}, and many of them could be associated with stellar streams. This motivates the hunt for dynamically cold stellar streams in the Galactic plane. The detection of stellar streams crossing the disk plane is accessible, especially, given the continued improvements in quality of the \emph{Gaia} data expected over the coming years, which will extend the detection horizon for stellar substructures. We envision applying our method to a suitable stellar stream in the near future.

\section*{Acknowledgements}

We would like to thank the anonymous referee for their contributions to this article.

We acknowledge support by the Oskar Klein Centre for Cosmoparticle Physics and Vetenskapsr{\aa}det (Swedish Research Council): AW through No. 621-2014-5772; KM, SS and PFdS through No. 638-2013-8993.

This research utilised the following open-source Python packages: \textsc{Isochrone} \citep{isochrones}, \textsc{Matplotlib} \citep{Hunter:2007}, \textsc{gala} \citep{gala}, \textsc{TensorFlow} \citep{tensorflow2015-whitepaper}.




\bibliographystyle{mnras}
\bibliography{example} 




\appendix

\section{Posterior probability density}\label{app:posterior}

The posterior, which is formulated in equation~\eqref{eq:posterior}, is a product of a prior probability of the model parameters, and $N$ six-dimensional integrals over a star's phase-space probability density and data likelihood, where $N$ is the number of observed stars in a stream. In order to make the posterior computationally tractable, the integrals over a star's phase-space coordinates can be analytically reduced, such that only a one-dimensional numerical integration over a star's distance remains. This is described in details below.

\subsection{Prior}\label{app:prior}

We use a wide flat box prior for the following parameters, with lower and upper bounds,
\begin{itemize}
    \item $\Sigma_{240} \in [0,200]~\Msunppcsquare$,
    \item $\Us_0, \Vs_0, \Ws_0 \in [-500,500]~\kmsec$,
    \item $\sigma_{\Us}, \sigma_{\Vs}, \sigma_{\Ws} \in [0,10]~\kmsec$,
    \item $\sigma_{\Xs}, \sigma_{\Ys} \in [0,100]~\pc$,
    \item $d_0 \in [0,5]~\kpc$,
    \item $l_0 \in [0,2\pi]$.
\end{itemize}
The following parameters have Gaussian priors centred on their true value (with standard deviations):
\begin{itemize}
    \item $F_{\Xs}$ (2.5 \% of its true value),
    \item $Z_{\odot}$ ($5~\pc$),
    \item $U_{\odot},V_{\odot}$ (5 km/s),
    \item $W_{\odot}$ (0.1 km/s).
\end{itemize}
See Section~\ref{sec:gal_model} for more details. The uncertainties are representative for how well these parameters are constrained for our own Sun and Galaxy \citep{10.1111/j.1365-2966.2010.16253.x,2019ApJ...871..120E}.

\subsection{Data likelihood}\label{app:likelihood}

Given a star with index $i$, with intrinsic position and velocity formulated in terms of the solar rest frame coordinates $\boldsymbol{X}_i$ and $\boldsymbol{V}_i$, the likelihood of its data is
\begin{equation}\label{eq:likelihood}
\begin{split}
    & \Pr(\data_i \, | \, \boldsymbol{X}_i, \boldsymbol{V}_i) \propto
    \delta[\hat{l}_i-l(\boldsymbol{X}_i)] \times \delta[\hat{b}_i-b(\boldsymbol{X}_i)] \\
    & \times \mathcal{N}[\hat{\mu}_{l,i} \,|\, \mu_l(\boldsymbol{X}_i,\boldsymbol{V}_i),\, \hat{\sigma}_{\mu,i}]
    \times \mathcal{N}[\hat{\mu}_{b,i} \,|\, \mu_b(\boldsymbol{X}_i,\boldsymbol{V}_i),\, \hat{\sigma}_{\mu,i}] \\
    & \times \mathcal{N}[\hat{\varpi}_i \,|\, \varpi(\boldsymbol{X}_i),\, \hat{\sigma}_{\varpi,i}],
\end{split}
\end{equation}
where the observables without hats ($l_i$, $b_i$, $\varpi_i$, $\mu_{l,i}$, $\mu_{b,i}$) are given directly by $\boldsymbol{X}_i$ and $\boldsymbol{V}_i$, assuming no observational errors (for example $\varpi(\boldsymbol{X}_i) = |\boldsymbol{X}_i|^{-1}~\kpc\times\mas$). The quantities with hats are the actual data observables, with associated uncertainties, as discussed in Section~\ref{sec:sim_streams}. The likelihood above is written in terms of solar coordinates, but can equally well be formulated in terms of $\boldsymbol{\Xs}_i$, $\boldsymbol{\Vs}_i$, and $\popp_\odot$. For the angular positions, observational uncertainties are neglected, which is why they are written in terms of Dirac delta functions. If radial velocity information is available, we add an additional factor $\mathcal{N}[\hat{v}_{\text{RV},i} \,|\, v_{\text{RV}}(\boldsymbol{X}_i,\boldsymbol{V}_i),\, \hat{\sigma}_{\text{RV},i}]$ to the likelihood.

\subsection{Coordinate system transformations}\label{app:transformations}

The subset of model parameters denoted $\popp_\odot$ describe the transformation from ``stream frame'' coordinates to solar coordinates. The position of the stream's mid-plane passage relative to the Sun is equal to
\begin{equation}
    \boldsymbol{X}_0 \equiv
    \begin{bmatrix}
        X_0 \\
        Y_0 \\
        Z_0 \\
    \end{bmatrix}
    =
    \begin{bmatrix}
        d_0\cos l_0  \\
        d_0\sin l_0 \\
        -Z_\odot \\
    \end{bmatrix}.
\end{equation}
The transformation from spatial coordinates $\boldsymbol{\Xs}$ to $\boldsymbol{X}$ is
\begin{equation}
    \begin{bmatrix}
        X \\
        Y \\
        Z \\
    \end{bmatrix}
    =
    \begin{bmatrix}
        \cos l_\text{diff.} & \sin l_\text{diff.} & 0 \\
        -\sin l_\text{diff.} & \cos l_\text{diff.} & 0 \\
        0 & 0 & 1 \\
    \end{bmatrix}\times
    \begin{bmatrix}
        \Xs-\Xs_0 \\
        \Ys-\Ys_0 \\
        \Zs \\
    \end{bmatrix}
    + \boldsymbol{X}_0,
\end{equation}
where $l_\text{diff.}$ corresponds to the angular difference between the direction of $X$ and $\Xs$ (the direction of the Galactic centre as seen from the Sun, and as seen from the stream's position). This angle is given by $d_0$, $l_0$, and the Galactic radius of the Sun $R_\odot$, according to the relation
\begin{equation}\label{eq:ldiff}
    \sin(l_\text{diff.}) = 
    \frac{d_0 \sin l_0}{\sqrt{d_0^2 + R_\odot^2 -2 d_0 R_\odot \cos l_0 }}.
\end{equation}

The transformation from velocities $\boldsymbol{\Vs}$ to $\boldsymbol{V}$ is
\begin{equation}
    \begin{bmatrix}
        U \\
        V \\
        W \\
    \end{bmatrix}
    =
    \begin{bmatrix}
        \cos l_\text{diff.} & \sin l_\text{diff.} & 0 \\
        -\sin l_\text{diff.} & \cos l_\text{diff.} & 0 \\
        0 & 0 & 1 \\
    \end{bmatrix}\times
    \begin{bmatrix}
        \Us \\
        \Vs \\
        \Ws \\
    \end{bmatrix}
    - \begin{bmatrix}
        U_\odot \\
        V_\odot \\
        W_\odot \\
    \end{bmatrix}.
\end{equation}

\subsection{Analytic reduction}\label{app:reduction}

The six-dimensional integrals can in large part be computed analytically. The angular coordinates ($\hat{l}$ and $\hat{b}$) have negligible uncertainties, such that only a numerical integral over the distance of a star with respect to the Sun is necessary. We denote this distance $s$. Focusing first on the integration over spatial positions for a star with index $i$, it can be rewritten as
\begin{equation}
\begin{split}
    \int &  \Pr(\data_i \, | \, \boldsymbol{\Xs}_i,\, \popp_\odot) \,
    f(\boldsymbol{\Xs}_i \, | \, \popp_S) \,
    \de^3 \boldsymbol{\Xs}_i = \\
    \int & \delta[\hat{l}_i-l(\boldsymbol{\Xs}_i,\popp_\odot)]\, \delta[\hat{b}_i-b(\boldsymbol{\Xs}_i,\popp_\odot)] \\ 
    & \times \mathcal{N}[\hat{\varpi}_i \,|\, \varpi(\boldsymbol{\Xs}_i,\popp_\odot), \sigma_{\varpi,i}]\,
    f(\boldsymbol{\Xs}_i \, | \, \popp_S) \,
    \de^3 \boldsymbol{\Xs}_i \propto \\
    \int & \mathcal{N}[\hat{\varpi}_i \,|\, \varpi(s), \sigma_{\varpi,i}]\,
    f[\boldsymbol{\Xs}_i(\hat{l}_i,\hat{b}_i,s,\popp_\odot) \, | \, \popp_S] \,
    s^2 \de s,
\end{split}
\end{equation}
where the intrinsic position of a star relative to the Sun is now completely fixed by $\hat{l}_i$, $\hat{b}_i$, and $s$. In this new formulation, a star's position in $\boldsymbol{\Xs}$ coordinates also depends on the relative position of the stream with respect to the Sun, which is encoded in $\popp_\odot$. The factor $s^2$ is part of the Jacobian of spherical coordinates, which appears when integrating the $\delta$ functions.

For the integral over velocity, the observational uncertainties and the stream's intrinsic dispersion both follow Gaussian distributions, and can therefore be computed analytically. For a fixed spatial position, the velocity part of the integral in equation~\eqref{eq:posterior} is equal to
\begin{equation}
\begin{split}
    \int & \Pr(\data_i \, | \, \boldsymbol{\Vs}_i,\, \popp_\odot)\, f[\boldsymbol{\Vs}_i \, | \, \Zs_i(\hat{l}_i,\hat{b}_i,s,\popp_\odot),\popp_S] \,
    \de^3 \boldsymbol{\Vs}_i = \\
    & \mathcal{M}[ \hat{\boldsymbol{\mu}}(\data_i) \,|\, \boldsymbol{\mu}(\hat{l}_i,\hat{b}_i,s,\popp), \boldsymbol{\Sigma}_\text{sum}(\data_i,s,\popp) ],
\end{split}
\end{equation}
where $\mathcal{M}$ is the multivariate normal distribution defined
\begin{equation}
    \mathcal{M}(\boldsymbol{x} \,|\, \bar{\boldsymbol{x}},\, \boldsymbol{\Sigma}) = \dfrac{\exp\left[
    -\dfrac{1}{2}(\boldsymbol{x}-\bar{\boldsymbol{x}})^\top \boldsymbol{\Sigma}^{-1}(\boldsymbol{x}-\bar{\boldsymbol{x}})
    \right]}{\sqrt{(2\pi)^q|\boldsymbol{\Sigma}|}},
\end{equation}
where $q$ is the dimension of $\boldsymbol{x}$. The quantity
\begin{equation}
    \hat{\boldsymbol{\mu}}(\data_i) =
    \begin{bmatrix}
    \hat{\mu}_{l,i} \\
    \hat{\mu}_{b,i} \\
    \hat{v}_{\text{RV},i}
    \end{bmatrix},
\end{equation}
is a vector with the observed proper motions and radial velocity. The quantity $\boldsymbol{\mu}$ is the corresponding vector as given by the ideal orbit of the stream model, which is defined
\begin{equation}\label{eq:velocity_trans}
\begin{split}
    & \boldsymbol{\mu}(\hat{l}_i,\hat{b}_i,s,\popp) = \\
    & \boldsymbol{M}(s)\times
    \boldsymbol{R}(\hat{l}_i+l_\text{diff.},\hat{b}_i) \times
   \begin{bmatrix}
    \Us[\Zs(\hat{l}_i,\hat{b}_i,s,\popp_\odot),\, \popp_S] \\
    \Vs[\Zs(\hat{l}_i,\hat{b}_i,s,\popp_\odot),\, \popp_S] \\
    \Ws[\Zs(\hat{l}_i,\hat{b}_i,s,\popp_\odot),\, \popp_S]
    \end{bmatrix},
\end{split}
\end{equation}
where $\Us[...]$, $\Vs[...]$, and $\Ws[...]$ are given by equation~\eqref{eq:vels_model}. The rotational matrix
\begin{equation}
	\boldsymbol{R}(l,b) =
    \begin{bmatrix}
    -\sin(l) &  \cos(l) & 0 \\
    -\cos(l)\sin(b) &  -\sin(l)\sin(b) & \cos(b)  \\
    \cos(l)\cos(b) &  \sin(l)\cos(b) & \sin(b)
    \end{bmatrix}
\end{equation}
transforms the solar coordinate system velocities ($U$, $V$, $W$) to velocities in the latitudinal, longitudinal, and radial directions ($v_l$, $v_b$, $v_{\text{RV}}$). The quantity $l_\text{diff.}$ corresponds to the angular difference between the direction of $X$ and $\Xs$, as defined in equation~\eqref{eq:ldiff}. The diagonal matrix
\begin{equation}
	\boldsymbol{M}(s) =
    \text{diag.}\bigg[ (k_\mu s)^{-1},\, (k_\mu s)^{-1},\, 1 \bigg],
\end{equation}
where $k_\mu = 4.74057 \, \kpc^{-1} \times \mas^{-1} \times \yr \times \kmsec$, corresponds to the transformation from velocities $v_l$ and $v_b$ to proper motions $\mu_l$ and $\mu_b$.

The covariance matrix $\boldsymbol{\Sigma}_\text{sum}$ is a sum of covariance matrices associated with observational uncertainties and the stream's intrinsic velocity dispersion around it's ideal orbit, like
\begin{equation}
    \boldsymbol{\Sigma}_\text{sum}(\data_i,s,\popp) =
    \boldsymbol{\Sigma}_{\hat{\mu}}(\data_i) +
    \boldsymbol{\Sigma}_S(\data_i,s,\popp).
\end{equation}
where
\begin{equation}
    \boldsymbol{\Sigma}_{\hat{\mu}}(\data_i) = 
    \begin{bmatrix}
    \hat{\sigma}_{\mu,i}^2 &  0 & 0 \\
    0 &  \hat{\sigma}_{\mu,i}^2 & 0  \\
    0 &  0 & \hat{\sigma}_{\text{RV},i}^2
    \end{bmatrix},
\end{equation}
and
\begin{equation}
\begin{split}
    & \boldsymbol{\Sigma}_S(\data_i,s,\popp) =
    \boldsymbol{M}(s) \times
    \boldsymbol{R}(\hat{l}_i+l_\text{diff.},\hat{b}_i) \\
    & \times
    \begin{bmatrix}
    \sigma_{\Vs}^2 &  0 & 0 \\
    0 &  \sigma_{\Us}^2 & 0  \\
    0 &  0 & \sigma_{\Ws}^2
    \end{bmatrix}
    \times
    \boldsymbol{R}^\top(\hat{l}_i+l_\text{diff.},\hat{b}_i)
    \times\boldsymbol{M}^\top(s).
\end{split}
\end{equation}
The matrices $\boldsymbol{M}$ and $\boldsymbol{R}$ are the same as in equation~\eqref{eq:velocity_trans}. Although we have not included error correlations or separate uncertainties for the two proper motions when testing our model on mock data, this is easily incorporated in $\boldsymbol{\Sigma}_{\hat{\mu}}(\data_i)$ when working with actual \emph{Gaia} data.

Using the simplifications described above, the posterior density becomes
\begin{equation}\label{eq:posterior_reduced}
\begin{split}
    & \pr(\popp \, | \, \data_{i=\{1,...,N\}}) = \\
    & \pr(\popp)\prod_{i=1}^N \int
    \mathcal{N}[\hat{\varpi}_i \,|\, \varpi(s),\, \hat{\sigma}_{\varpi,i}] \times
    f[\boldsymbol{\Xs}_i(\hat{l}_i,\hat{b}_i,s,\popp_\odot) \, | \, \popp_S]  \\
    & \times
    \mathcal{M}[ \hat{\boldsymbol{\mu}}(\data_i) \,|\, \boldsymbol{\mu}(\hat{l}_i,\hat{b}_i,s,\popp), \boldsymbol{\Sigma}_\text{sum}(\data_i,s,\popp) ] \,
    s^2 \de s,
\end{split}
\end{equation}
now containing $N$ one-dimensional integrals, which is significantly less intensive computationally.

\subsection{MCMC sampling}\label{app:sampling}

The model is implemented in \texttt{TensorFlow}, and the posterior probability density is explored using Hamiltonian Monte-Carlo (HMC). The HMC algorithm utilises differentiation of the posterior density function with respect to its free parameters, which allows for efficient sampling despite the high number of degrees of freedom.

The posterior probability density, as written in equation~\eqref{eq:posterior_reduced}, contains $N$ separable integrals over distance $s$, one for each star in our stellar stream sample. These integrals are somewhat expensive to compute, but inference can be significantly sped up by promoting the distance of each object to a free nuisance parameter of the posterior. We rewrite the posterior density to read
\begin{equation}\label{eq:posterior_reduced_expanded}
\begin{split}
    & \pr(\popp,s_{i=\{1,...,N\}} \, | \, \data_{i=\{1,...,N\}}) = \\
    & \pr(\popp)\prod_{i=1}^N
    \mathcal{N}[\hat{\varpi}_i \,|\, \varpi(s_i),\, \hat{\sigma}_{\varpi,i}] \times
    f[\boldsymbol{\Xs}_i(\hat{l}_i,\hat{b}_i,s_i,\popp_\odot) \, | \, \popp_S]  \\
    & \times
    \mathcal{M}[ \hat{\boldsymbol{\mu}}(\data_i) \,|\, \boldsymbol{\mu}(\hat{l}_i,\hat{b}_i,s_i,\popp), \boldsymbol{\Sigma}_\text{sum}(\data_i,s_i,\popp) ] \,
    s_i^2,
\end{split}
\end{equation}
now having a total of $16+N$ free parameters. Sampling this expanded posterior density function effectively marginalises over the $N$ distance parameters $s_{i=\{1,...,N\}}$, which is equivalent to the integral formulation of equation~\eqref{eq:posterior_reduced}.

The HMC run is started with a thorough burn-in phase, where the mode of the posterior density is located, and the step-size is tuned (this is done by tuning the HMC mass matrix, which is assumed to be diagonal). The HMC is then run for $10^6$ steps and thinned by a factor $10^2$, giving a chain of $10^4$ posterior realisations. We test convergence of the HMC chains by auto-correlation and see that they are well sampled, where (at the very least) every ten steps of the HMC is an independently drawn realisation of the posterior distribution. We also see that the final run of the HMC gives a similar result with respect to the last burn-in phase.

\section{Stream phase-space coordinates}\label{app:coords}

In Fig.~\ref{fig:coords_S1}--\ref{fig:coords_S4} below, we show the intrinsic phase-space coordinates of streams S1--S4. The $\Xs$ and $\Ys$ values are normalised to their median values in these figures.

\begin{figure*}
    \centering
	\includegraphics[width=2.0\columnwidth]{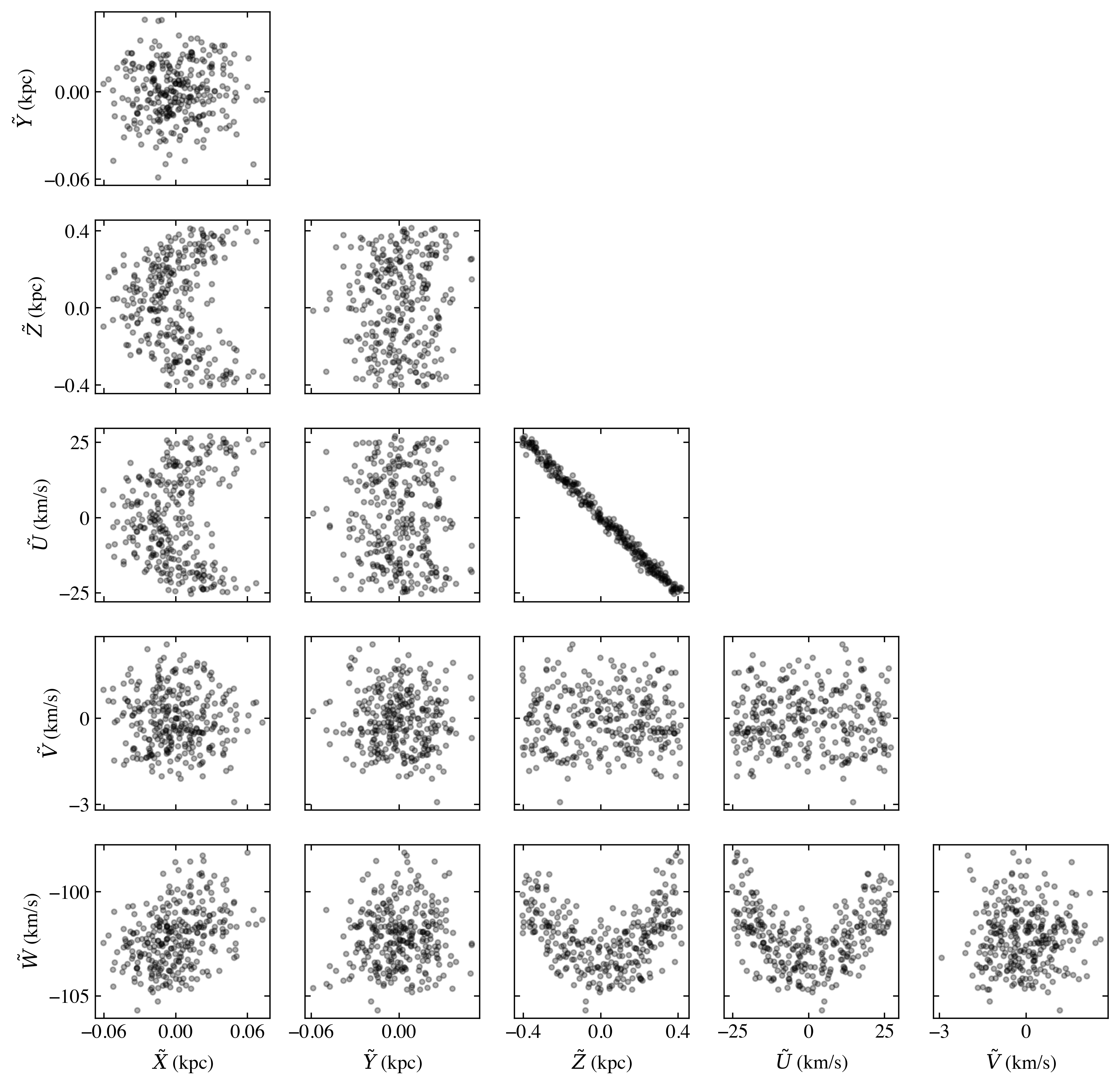}
    \caption{Intrinsic phase-space coordinates of stream S1.}
    \label{fig:coords_S1}
\end{figure*}

\begin{figure*}
    \centering
	\includegraphics[width=2.\columnwidth]{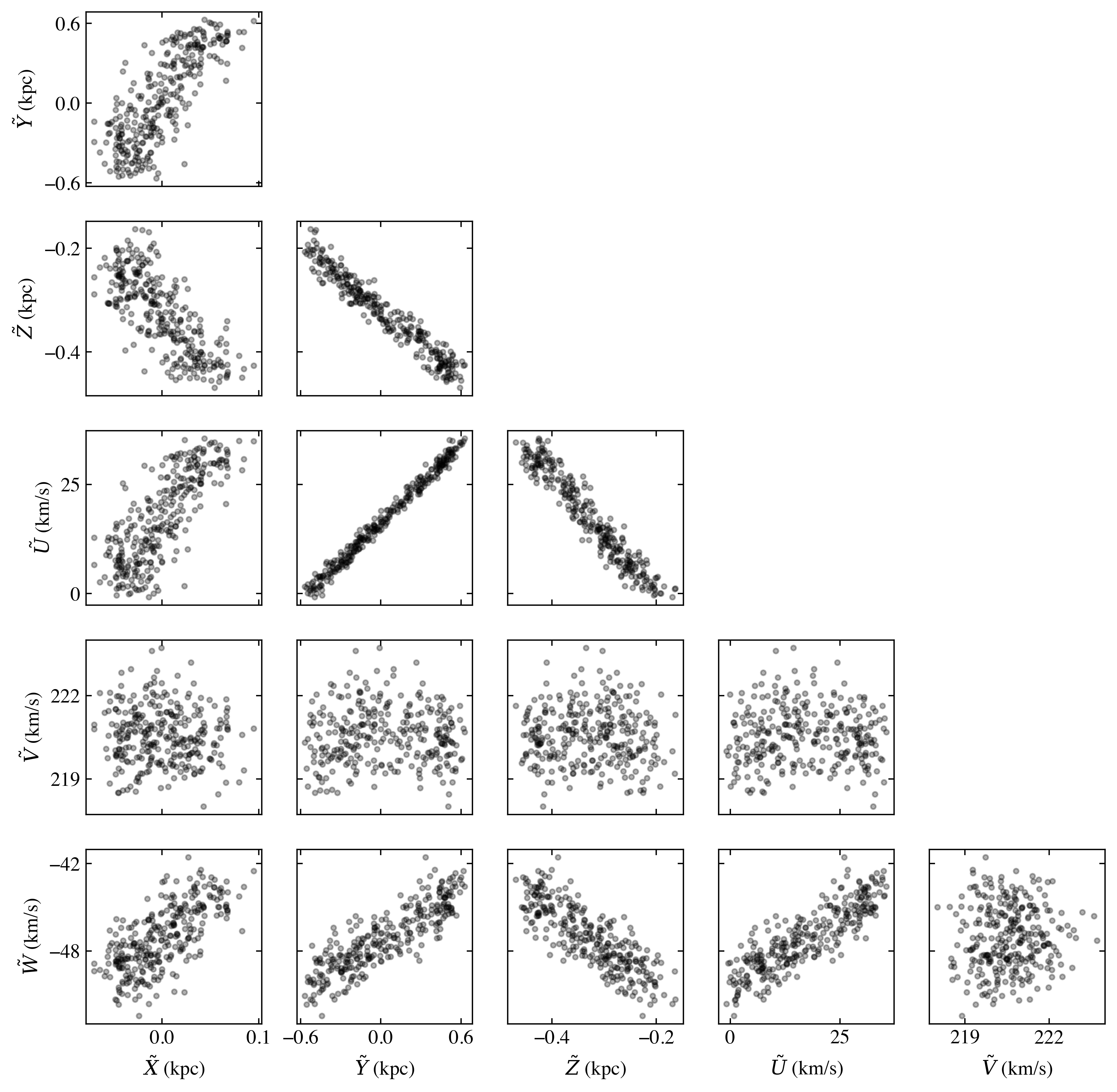}
    \caption{Intrinsic phase-space coordinates of stream S2.}
    \label{fig:coords_S2}
\end{figure*}

\begin{figure*}
    \centering
	\includegraphics[width=2.\columnwidth]{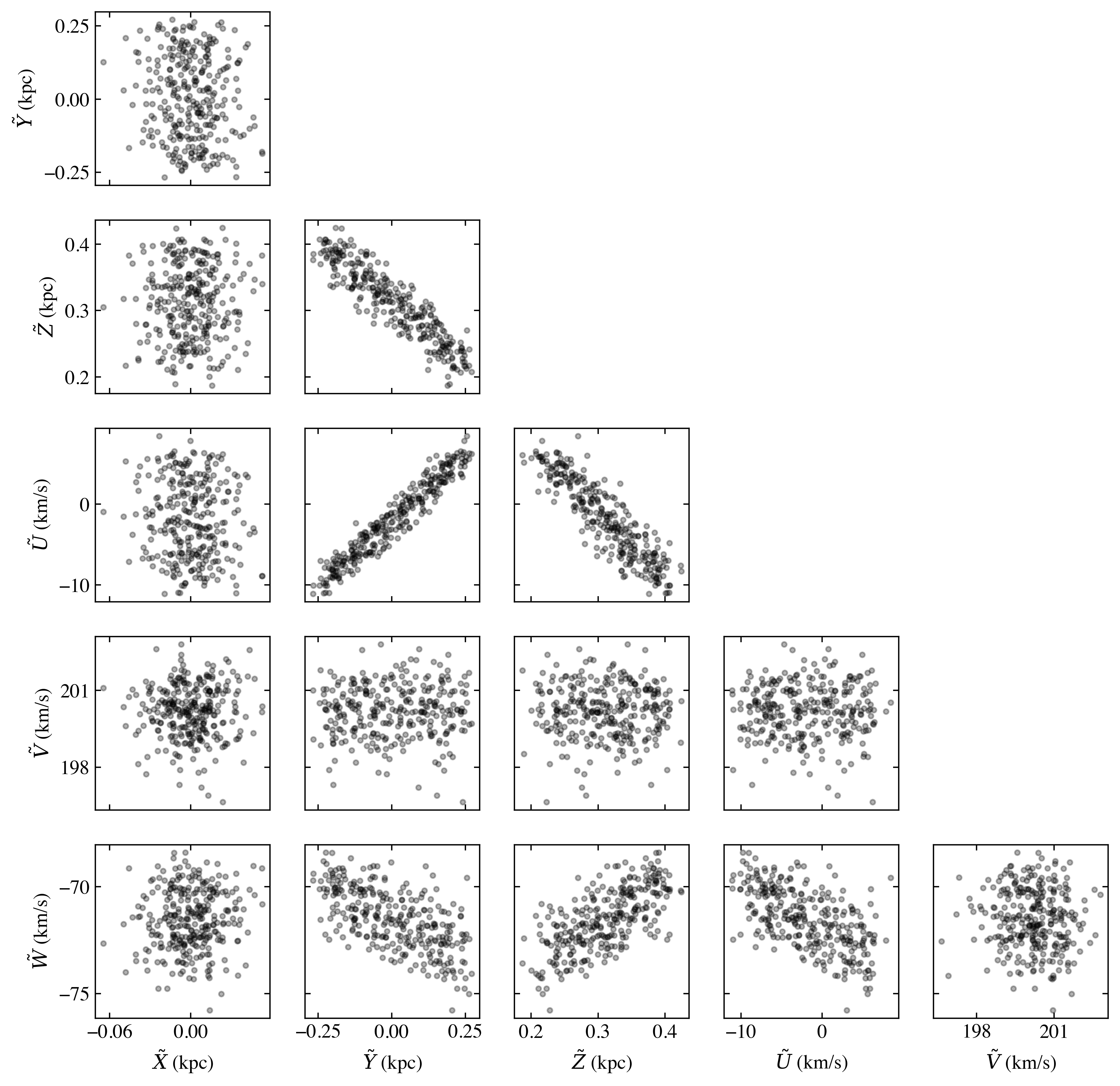}
    \caption{Intrinsic phase-space coordinates of stream S3.}
    \label{fig:coords_S3}
\end{figure*}

\begin{figure*}
    \centering
	\includegraphics[width=2.\columnwidth]{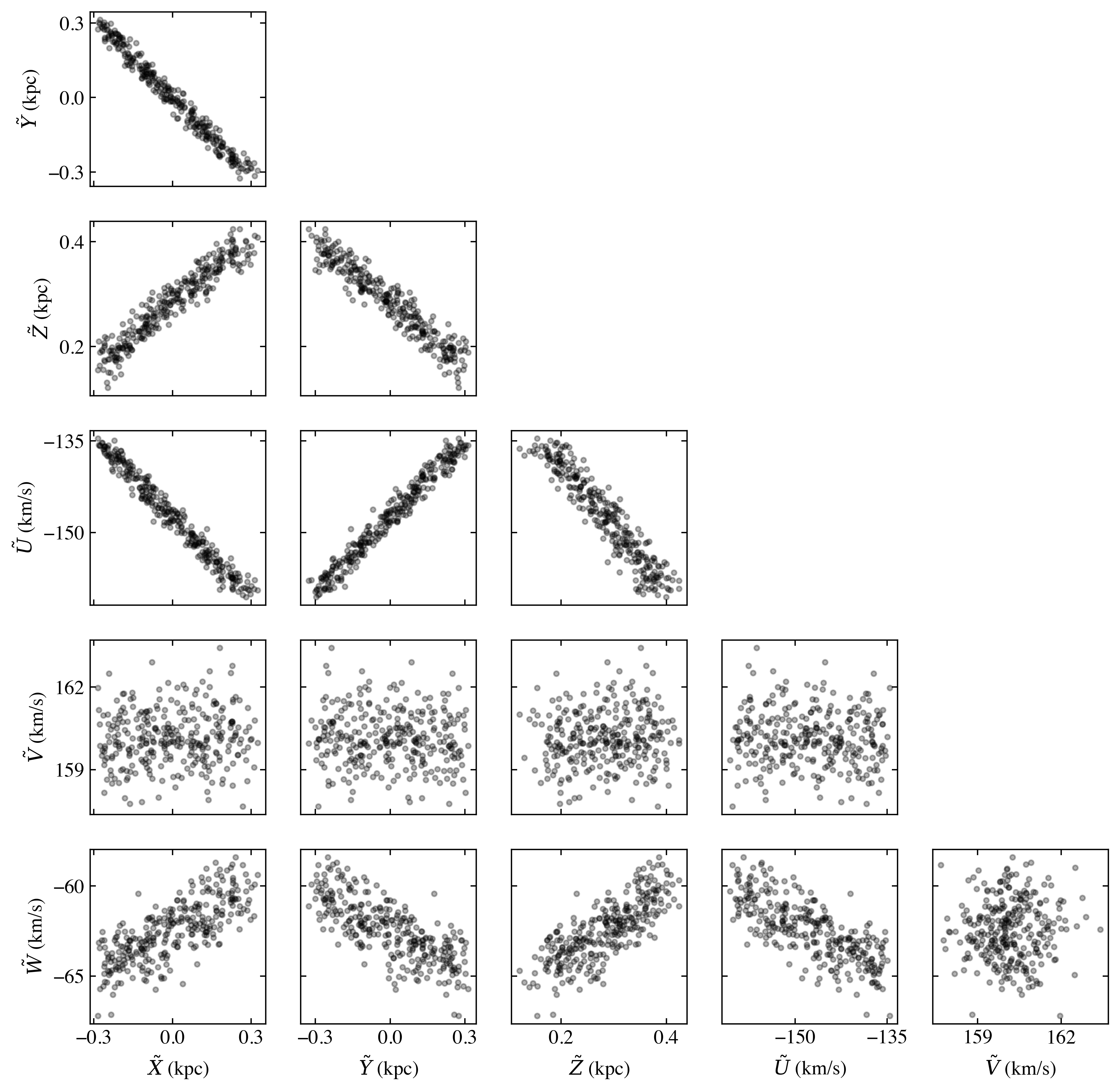}
    \caption{Intrinsic phase-space coordinates of stream S4.}
    \label{fig:coords_S4}
\end{figure*}


\bsp	
\label{lastpage}
\end{document}